\documentclass[aps,pra,twocolumn,reprint,superscriptaddress]{revtex4-1}
\usepackage{graphicx}  
\usepackage{dcolumn}   
\usepackage{bm}        
\usepackage{amssymb}   % for math
\usepackage{physics}
\usepackage{tikz}
\usepackage{gensymb}
\usetikzlibrary{shapes,snakes}
\usetikzlibrary{shapes.geometric}
\usepackage{pgfplots}
\usepackage{chngcntr}
\usepackage{dsfont}
\usepackage[caption = false]{subfig}
\usepackage{hyperref}
\usepackage{xcolor}
\usepackage{float}
\usepackage{cleveref}
\usepackage[utf8]{inputenc}
\usepackage[english]{babel}
 \usepackage{amsthm}

\theoremstyle{definition}

\newcommand{\newc}{\newcommand}
\newc{\beq}{\begin{equation}}
\newc{\eeq}{\end{equation}}
\newc{\kt}{\rangle}
\newc{\br}{\langle}
\newc{\beqa}{\begin{eqnarray}}
\newc{\eeqa}{\end{eqnarray}}
\newc{\pr}{\prime}
\newc{\longra}{\longrightarrow}
\newc{\ot}{\otimes}
\newc{\rarrow}{\rightarrow}
\newc{\h}{\hat}
\newc{\bom}{\boldmath}
\newc{\btd}{\bigtriangledown}
\newc{\al}{\alpha_k}
\newc{\be}{\beta_k}
\newc{\ld}{\lambda}
\newc{\sg}{\sigma}
\newc{\p}{\psi}
\newc{\eps}{\epsilon}
\newc{\om}{\omega}
\newc{\mb}{\mbox}
\newc{\tm}{\times}
\newc{\ra}{\rightarrow}
\newc{\non}{\nonumber}
\newc{\ul}{\underline}
\newc{\hs}{\hspace}
\newc{\longla}{\longleftarrow}
\newc{\ts}{\textstyle}
\newc{\f}{\frac}
\newc{\df}{\dfrac}
\newc{\ovl}{\overline}
\newc{\bc}{\begin{center}}
\newc{\ec}{\end{center}}
\newc{\dg}{\dagger}
\newc{\T}{\mathcal{U}}
\newc{\Tp}{\mathcal{V}}
\newc{\J}{\mathsf{J}}
\newc{\sfL}{\mathsf{L}}
\newc{\C}{\mathsf{C}}
\newc{\B}{\mathsf{M}}
\newc{\V}{\mathsf{V}}
\newc{\red}{\textcolor{red}}

\begin{document}
% The following information is for internal review, please remove them for submission
\widetext

\title{Quantum walks with quantum chaotic coins: Of the Loschmidt echo, classical limit and thermalization}
      % D0 authors (remove the first 3 lines
                             % of this file prior to submission, they
                             % contain a time stamp for the authorlist)
                             % (includes institutions and visitors)

\author{Sivaprasad Omanakuttan}
\email[]{somanakuttan@unm.edu}
\affiliation{Center for Quantum Information and Control (CQuIC),Department of Physics and Astronomy, University of New Mexico, Albuquerque, New Mexico 87131, USA}
\author{Arul Lakshminarayan}
\email[]{arul@physics.iitm.ac.in}
%\thanks{}
\affiliation{Department of Physics, Indian Institute of Technology Madras, Chennai 600036, India}

\date{\today}

\begin{abstract}

Coined discrete-time quantum walks are studied using simple deterministic dynamical systems as coins whose classical limit can range from being integrable to chaotic. It is shown that a Loschmidt echo like fidelity plays a central role and when the coin is chaotic this is approximately the characteristic function of a classical random walker. Thus the classical binomial distribution arises as a limit of the quantum walk and 
the walker exhibits diffusive growth before eventually becoming ballistic. The coin-walker entanglement growth is shown to be logarithmic in time as in the case of many-body localization and coupled kicked rotors, and saturates to a value that depends on the relative coin and walker space dimensions. In a coin dominated scenario, the chaos can thermalize the quantum walk to typical random states such that the entanglement saturates
at the Haar averaged Page value, unlike in a walker dominated case when atypical states seem to be produced.
 \end{abstract}
\maketitle

\section{Introduction}
In the current era of ``noisy intermediate-scale quantum" (NISQ) technologies \cite{preskill2018quantum} the quest for quantum supremacy is heating up \cite{arute2019quantum}, although a general purpose quantum computer which outperforms a state-of-the-art supercomputer remains an, apparently, distant goal. The most significant challenge is to minimize environment effects in order to harness the possible ``quantum advantages''. Apart from such external noise, decoherence-like effects
can take place when part of the system is non-integrable or chaotic \cite{PhysRevE.62.3504,frahm2004quantum}. We consider this question in the context of quantum walks which is considered as one of the platforms for doing universal quantum computation \cite{childs2009universal,lovett2010universal} and quantum search algorithms \cite{childs2004spatial,aaronson2003quantum}.

The discrete-time coined quantum walk \cite{aharonov1993quantum,kempe2003quantum,nayak2000quantum}, a quantum version of the classical random walker paradigm of diffusion is also relevant to quantum transport, and many other fields of physics \cite{chandrashekar2008optimizing,zhang2016creating,
di2014quantum,genske2013electric,muraleedharan2019quantum}. The system is a bipartite one, consisting of a ``coin" and a ``walker". The coin states span an $M$-dimensional Hilbert space and the coin toss is an unitary ``flip" of the coin that transforms states in this space. The walker is an $N$ dimensional Hilbert space consisting of $N$ sites on which the dynamics can only be a translation by one site. The interaction between the coin and the 
walker is determined by a controlled-translation of the walker, controlled by if the projector of the coin in some fixed basis is in a set identified
as ``Head" or ``Tail". 
%This is followed by a coin toss, that is the unitary coin dynamics and then again the walker chooses to perform the controlled-translation. This protocol is a purely coherent one involving no measurements of the coin state and the various possibilities add to give the quantum walk its uniqueness.}

\begin{figure}[]
%\subfloat[$M\gg N$]{\includegraphics[width =1.8in]{M_largerthan_N}}
%\subfloat[$M\gg N$]{\includegraphics[width =1.8in]{entanglement_33}}
%\subfloat[$M<<N, M<30$]{\includegraphics[width =2.7in]{N_larger_than_M}}
\includegraphics[width=0.48\textwidth]{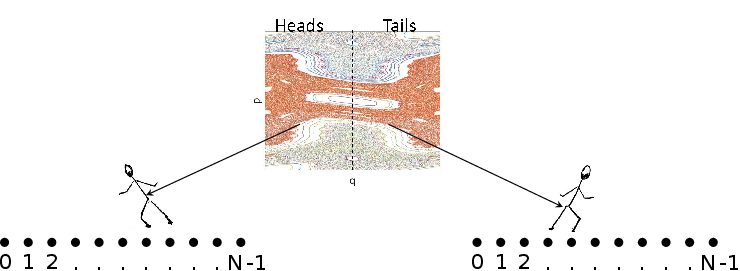}
%\subfloat[Entanglement]{\includegraphics[width =1.8in]{compare_variance_entanglement}}
\caption{Schematic of the quantum chaotic walk. The translation of the walker in a lattice depends on a coin which is the quantization of a classically choatic dynamical system. The walker moves left or right depending on whether the coin state is one of two possible orthonormal projectors on its Hilbert space. The projectors add to the whole space and their classical limits are the right or left of the dotted lines in the classical phase space respectively.}  \label{fig:classical_random_walk}
\end{figure}

 However, compared to its classical counterpart the coin in the case of the  quantum walk is reversible and unitary and leads to the coherent superpositions that ultimately leads to the quantum advantage of ballistic growth. It is also typically a two-state system that reflects the binary choices on a coin. Being devoid of random elements it is referred to simply as ``quantum walks". The classical random walker limit is realized from the quantum by the following drastic modifications: (i) measurement of the discrete quantum walk at each time step or (ii) the use of multiple independent coins, one at each time step, or (iii) by the explicit presence of decoherence that destroys quantum effects. It has also been shown that as long as only one coin is used, however large the coin's internal space (the dimension $M$) may be, asymptotically the quantum walk is ballistic \cite{PhysRevLett.91.130602}.

This leaves open the possibility that there are deterministic, reversible, unitary coins, such that there is diffusive classical growth
for some time that gives way eventually to a quantum ballistic growth due to the dominance of quantum effects. If the internal coin dynamics is 
non-integrable or quantum chaotic, we may expect that this is reflected in the nature of the quantum walk, just as a classically random coin 
is assumed in the classical walks. Deviations from randomness can alter the nature of the classical walk, and indeed the quantum coin
also has the power to alter the quantum walk.
A quantum chaotic coin leads to quantum walker diffusion over a finite time that can diverge with the coin size.
This is in accordance with standard classical-quantum correspondences of dynamical systems and allows for deterministic chaos to play the role of the stochastic coin in the classical random walk. Schematic of a random walk determined by a chaotic dynamical system is shown in Fig.~(\ref{fig:classical_random_walk}).

%\begin{figure}[]
%%\subfloat[$M\gg N$]{\includegraphics[width =1.8in]{M_largerthan_N}}
%%\subfloat[$M\gg N$]{\includegraphics[width =1.8in]{entanglement_33}}
%%\subfloat[$M<<N, M<30$]{\includegraphics[width =2.7in]{N_larger_than_M}}
%\includegraphics[scale=.32]{overall_picture}
%%\subfloat[Entanglement]{\includegraphics[width =1.8in]{compare_variance_entanglement}}
%\caption{\red{Schematic of the influence of a chaotic or regular coin dynamics  on a quantum walker. The regular coin subsystem recovers the probability distribution of the quantum walker with a ballistic spread. The quantum chaotic coin recovers the classical Gaussian ditribution of walker positions for a time that is proportional to the coin dimension.}}  \label{fig:choatic_vs_regular}
%\end{figure} 

Previous work on such walks includes a numerical study of quantum chaotic walks and its interpretation as the quantization of a classical transport model on a lattice \cite{lakshminarayan2003random}, while chaotic environments of a binary coin have been considered in \cite{ErmanPazSaraceno, Wojcik}, which are effectively identical models. In \cite{Wojcik} transport properties were studied and reviewed including an analytical demonstration of a diffusive growth till the Heisenberg time of the coin $ \sim M$.

Going beyond the second moment of the walker, for the case of quantum chaotic walks, we show analytically the emergence of the ubiquitous Gaussian walker distribution and hence normal diffusion by exploiting a somewhat surprising connection to the phenomenon of Loschmidt echo: the decay of fidelity upon forward and backward evolutions under slightly different Hamiltonians, a well-known diagnostic of quantum chaos and a measure of hypersensitivity of the dynamics. Using the Fermi-golden rule regime of a Loschmidt echo calculation of the coin dynamics, wherein the walker's conserved momenta play the role of parameter variation in forward and  backward time evolutions, we see the classical diffusive regime emerge. To reiterate, this happens in the absence of measurement, decoherence or multiple coins.

We study coin-walker entanglement which grows as $\sim \log(\sqrt{t})$ with a coefficient that is weakly changing with time, before saturating for finite lattices. The saturation value indicates thermalization if the walker or coin entropies approach the so-called Page value of random bipartite states.
This is a thermalization in the sense that the combined walker-coin state is as if it were chosen from a uniform distribution of pure states,
the Haar measure. This happens when the dimension of the coin dominates or is comparable to the walker space, in which case
the quantum chaos of the coin pervades the walker space leading to eventual thermalization such that the entanglement saturates
at the Haar average value, the Page value. In the walker dominated case the entanglement is smaller than the Page value, indicating some sort of 
localization and is reflected in the density of the spectra of the reduced density matrices not thermalizing to the Marchenko-Pastur law. 
Even-odd lattice size effects have dramatic consequences, as this saturation value is over the full rank of the coin space or half of it, depending on if the lattice is bipartite or not. It is also pointed out that this is seen most transparently in a path-integral form of the quantum walk.

\section{Setting of the quantum chaotic walk}

\subsection{The quantum walk and translational symmetry}

The quantum walk discussed here is on a $N$ node cyclic graph that is simply a linear lattice with periodic boundary conditions and is defined at any discrete time step by the unitary operator
\beq
\label{eq:qwU}
U=(P_R\otimes \T+P_L \otimes \T^{\dagger})(U_C \otimes \mathds{1}_N),
\eeq
 where $U_C$ is a coin toss operator in dimension $M$. Thus the quantum walk space is the tensor product $MN$ dimensional space. The projection operators $P_R$ and $P_L$ on the coin space are orthogonal and complementary: $P_R+P_L= \mathds{1}_M$ and $\T$ is the position translation operator which shifts the walker from one lattice site to the adjacent one, namely $\T |n \kt =|n+1\kt$ \cite{kempe2003quantum}. The coin's bias can be set by $\tr P_R/M= 1-\tr P_L/M$, and the walker transits to the left or right depending on the coin state's overlap with these projectors. In the following, we consider an unbiased walk and operate in a basis in which the projectors are diagonal: $P_L=\sum_{\alpha=0}^{M/2-1}|\alpha\kt\br\alpha|$ and  $P_R=\sum_{\alpha=M/2}^{M-1}|\alpha\kt\br\alpha|$, and we assume that $M$ is an even integer. Note that in this case we can consider the coin space to be the tensor product of a qubit and a $M/2$ dimensional subsystem, and the projectors $P_R$ and $P_L$ correspond to $|1\kt \br 1|\otimes I_{M/2}$ and $|0\kt \br 0|\otimes I_{M/2}$, hence the model may also be thought of as the walker with a two dimensional coin that is interacting with a larger system, a point of view adopted in \cite{ErmanPazSaraceno}.

The operator $U$ can be block diagonalized in the momentum basis of the walker in which $\T$ is diagonal: $\T |\tilde{k} \kt = e^{-2 \pi i k/N}|\tilde{k} \kt$, and $\br n |\tilde{k}\kt = e^{-2 \pi i kn/N}/\sqrt{N}$.
It follows that $U=\oplus_{k=0}^{N-1} U_k$, where 
\beq
\label{eq:UK}
U_k=(e^{-2 \pi i k/N} P_R+e^{2 \pi i k/N} P_L) U_C,
\eeq
or more explicitly the matrix elements $\br \alpha |U_k|\beta \kt$ 
\beq
 \label{eq:block form of unitary operaror}
 =\begin{cases}
 \exp(\frac{2\pi i k}{N})\bra{\alpha}U_C\ket{\beta}; 0\leq\alpha<\frac{M}{2}\\
  \exp(\frac{-2\pi i k}{N})\bra{\alpha}U_C\ket{\beta}; \frac{M}{2}\leq\alpha<M.
 \end{cases}
 \end{equation}

Consider the initial state $\ket{\psi(t=0)}=\ket{0}_C\ket{0}_W$, where $|0\kt_C$ is the zero momentum
state in the coin space and $|0\kt_W$ is the walker starting from the $0$ lattice site.
The state of the whole system after a time $t$ is then, 
\begin{equation}
%\begin{aligned}
\ket{\psi(t)}=
%&\frac{1}{\sqrt{N}}\bigoplus_{k=0}^{N-1}U_k^{t}\ket{0}_c\sum_{k=0}^{N-1}|\tilde{k}\kt\\
\frac{1}{\sqrt{N}}\sum_{k=0}^{N-1}U_k^t\ket{0}_C |\tilde{k}\kt,
%\end{aligned}
\end{equation}
with the pure-state density matrix $\rho(t)=|\psi(t)\kt \br \psi(t)|$. 
%\begin{equation}
%\begin{aligned}
%\rho(t)=&\frac{1}{N}\sum_{k,l=0}^{N-1}U_k^{t}\ket{0}\ket{\tilde{k}}\bra{\tilde{l}}\bra{0}U_l^{-t},
%\end{aligned}
%\label{eq:RD 1}
%\end{equation}
The coin state is the reduced density matrix, 
\begin{equation}
\rho_C(t)= \tr_W(\rho(t))=\frac{1}{N}\sum_{k=0}^{N-1}U_k^{t}\ket{0}_C \bra{0}U_{k}^{-t},
\label{eq:RCW}
\end{equation}
which is explicitly in the  Kraus-Sudarshan decomposed form of a quantum operation or channel \cite{kraus1983states, sudarshan1961stochastic}. It implies that the
coin state is decohered from the initial pure state as if it passes through
a channel with the Kraus or noise operators $\{ U_k^t/\sqrt{N},\, 0\leq k \leq N-1\} $.

The walker state is the reduced density matrix,
\begin{equation}
\begin{aligned}
\rho_W(t)=&\frac{1}{N}\sum_{k,l=0}^{N-1}  \bra{0} U_l^{-t} U_k^{t}\ket{0} |\tilde{k}\kt \br \tilde{l}|,\\
\end{aligned}
\label{eq:RCW 1}
\end{equation}
an $N\cross N$ matrix with elements, $ [\rho_W(t)]_{kl}=\frac{1}{N}\bra{0}U_l^{-t}U_k^{t}\ket{0}$. 
We have dropped the coin and walker subscripts from the states and unless explicitly specified $|0\kt$ will refer
to the zero momentum initial state of the coin.
The walker reduced density matrix in the site or position basis $\ket{n}$ is obtained by appropriate 
Fourier transforms:
\begin{equation}
\bra{n}\rho_W(t)\ket{n'}=\frac{1}{N^2}\sum_{k,l}\bra{0}U_l^{-t}U_k^{t}\ket{0}\exp[\frac{2\pi i (n'l-nk)}{N}].
\label{eq:RCW 2}
\end{equation}

Consider for $U_C$ a unitary operator with a well defined classical limit which has an integrable to chaotic transition as a function of a parameter. We study the change in the quantum walk based on this parameter to understand the effects when the coin achieves the ``deterministic'' randomness limit from an integrable one.

\subsection{The classical quantum Harper map as a coin}
While earlier studies in \cite{ErmanPazSaraceno, Wojcik} used the quantum baker's map which lacks such a parameter and is known to have non-generic features, we choose the Harper map, whose classical limit is 
\beq
\label{eq:Hmap}
\begin{split}
q_{t+1}&=q_t-\tau \sin(2 \pi p_t)\\
p_{t+1}&=p_t+\tau g\sin(2 \pi q_{t+1})
\end{split}
\eeq
and $(q,p)$ is on a unit torus and hence modulo-1 operation is assumed. This two-parameter area-preserving map is the Floquet map of the time-periodic Hamiltonian 
\begin{figure*}[]
\centering
\subfloat{\includegraphics[width =2.2in]{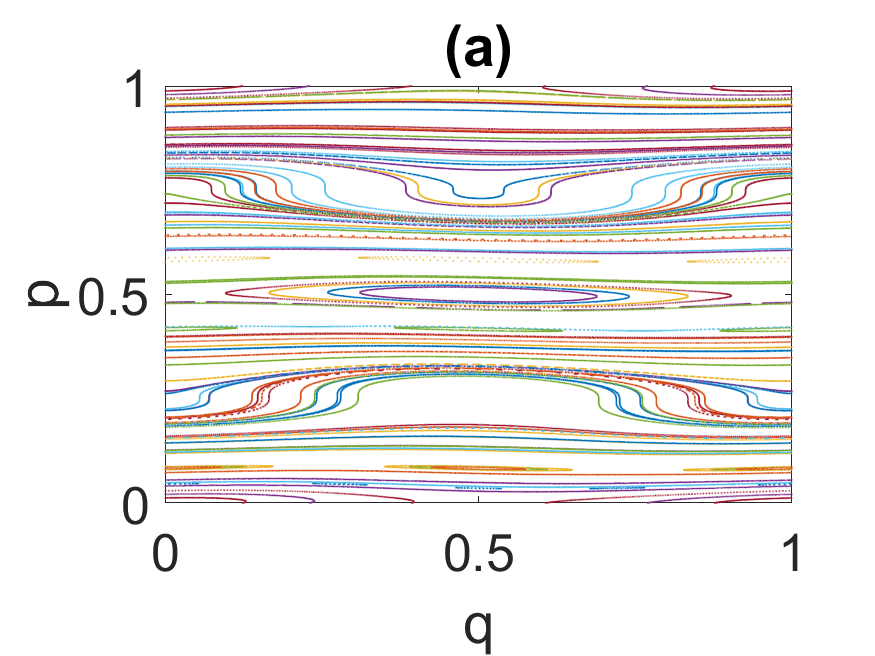}}
\subfloat{\includegraphics[width =2.2in]{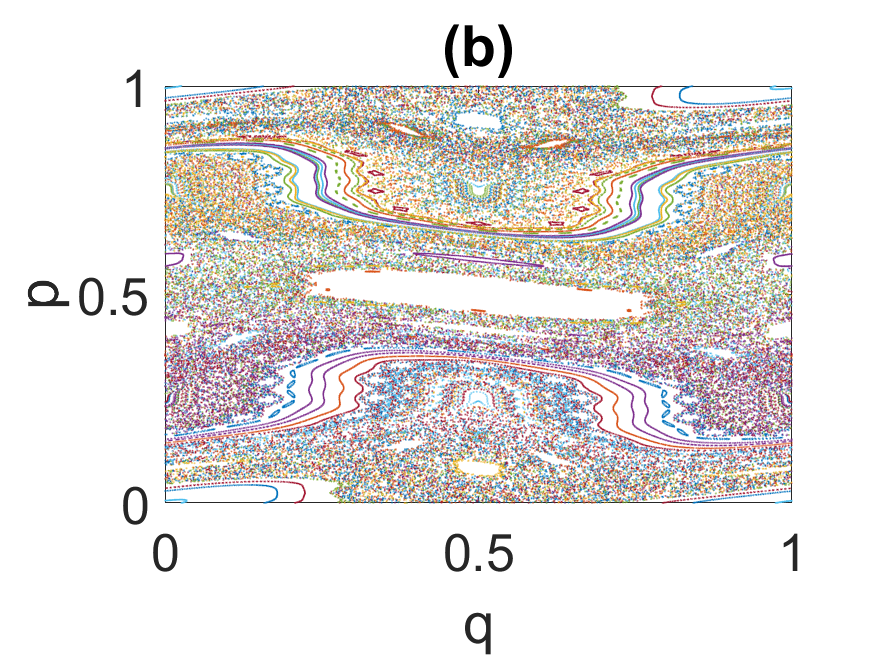}}
\subfloat{\includegraphics[width =2.2in]{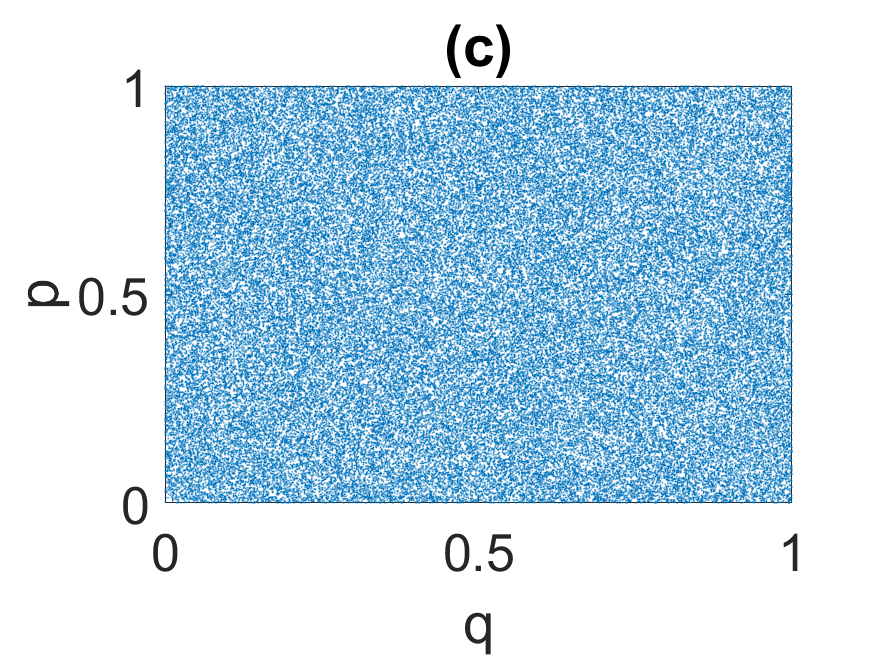}}

%\subfloat[g=0.01]{\includegraphics[width =2in]{h001}}
%\subfloat[g=0.05]{\includegraphics[width =2in]{h005}}
%\subfloat[g=0.1]{\includegraphics[width =2in]{h01}}\\
%\subfloat[g=0.5]{\includegraphics[width =2in]{h05}}
%\subfloat[g=1.0]{\includegraphics[width =2in]{harper1}}
%\subfloat[g=2.0]{\includegraphics[width =2in]{harper2}}
   \caption{Phase-space orbits of the kicked Harper map in Eq.~(\ref{eq:Hmap}) for $\tau=1$ and three representative values of $g$. For $g=0.01$ (a)  the dynamics is near integrable, while for $g=0.05$ (b) the dynamics is mixed with both chaotic and regular orbits, while for $g=0.4$ (c) there are no visible stable islands indicating full chaos. In the first two cases, $100$ random initial conditions have been iterated a $1000$ times, while in the last case only one random initial condition has been iterated $10^5$ times.  }\label{fig:harper}
\end{figure*}
\beq
\label{eq:HH}
H=\cos(2\pi p)+ g \cos (2 \pi q )\sum_{n=-\infty}^{\infty} \delta(2 \pi t /\tau-n)
\eeq
  connecting states just after consecutive kicks.
We choose to change $g$, keeping the time scale $\tau=1$. When $g=0$ the dynamics conserves momentum and is integrable, while for $g \neq 0$
it is non-integrable, and for $g \geq 0.05$, the dynamics of the system is largely chaotic and some phase space portraits are shown in Fig.(\ref{fig:harper}). Our choice of the Harper map over the baker is that the baker is fully chaotic and has special quantal features,
while the Harper is capable of showing a range of dynamics from integrable through mixed to fully chaotic and its quantization has
generic features.

The quantization of the kicked Harper model is given by the Floquet operator  
\beq
\label{eq:Qmap}
U_C=\exp[-i \frac{\tau\, g}{h} \cos(2\pi \hat{q})] \exp [-i \frac{\tau}{h} \cos(2\pi \hat{p})]. 
\eeq
Due to the unit torus classical phase space, the value of the scaled Planck constant is $h=1/M$, where $M$ is an integer and is the dimensionality of the coin space. There is a lattice of momentum and position states with values given by multiples of $h$. Thus in the momentum basis, with $0 \leq m,m' \leq M-1$, $U_C$ is
\begin{widetext}
\beq
\begin{aligned} 
\br m |U_C(g)|m' \kt= \frac{1}{M}\exp\left[-i \tau M \cos\left(\frac{2 \pi m'}{M} \right)\right] \,\sum_{k=0}^{M-1}\exp\left[-i \tau g M \cos\left(\frac{2 \pi k}{M}\right) \right] e^{\left[2\pi i k(m'-m)/M\right]}.
\end{aligned}
\eeq
\end{widetext}

\begin{figure*}
\subfloat{\includegraphics[width =2in]{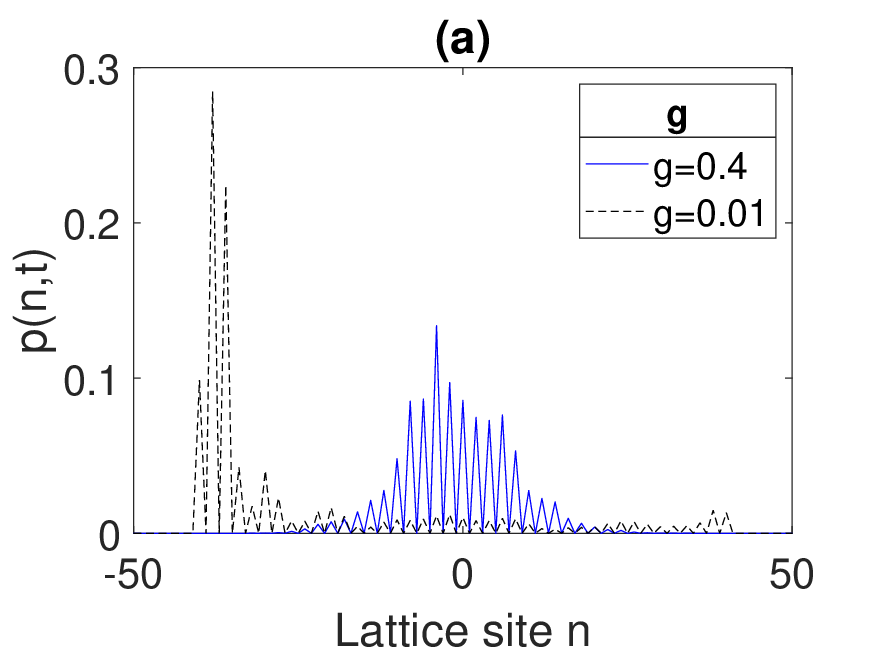}\label{fig:harper_1 a}}
\subfloat{\includegraphics[width =2in]{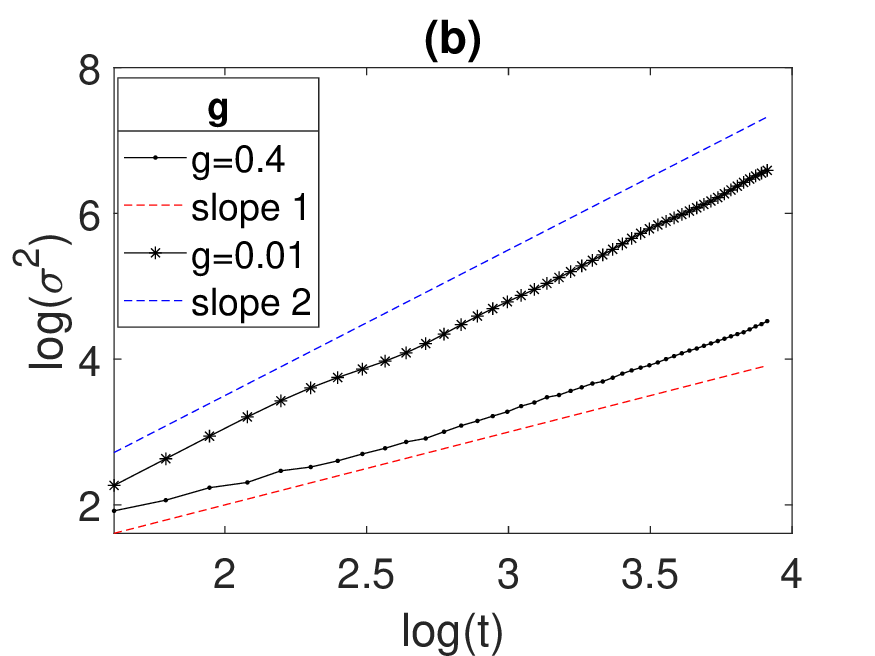} \label{fig:harper_1 b}}
\subfloat{\includegraphics[width =2in]{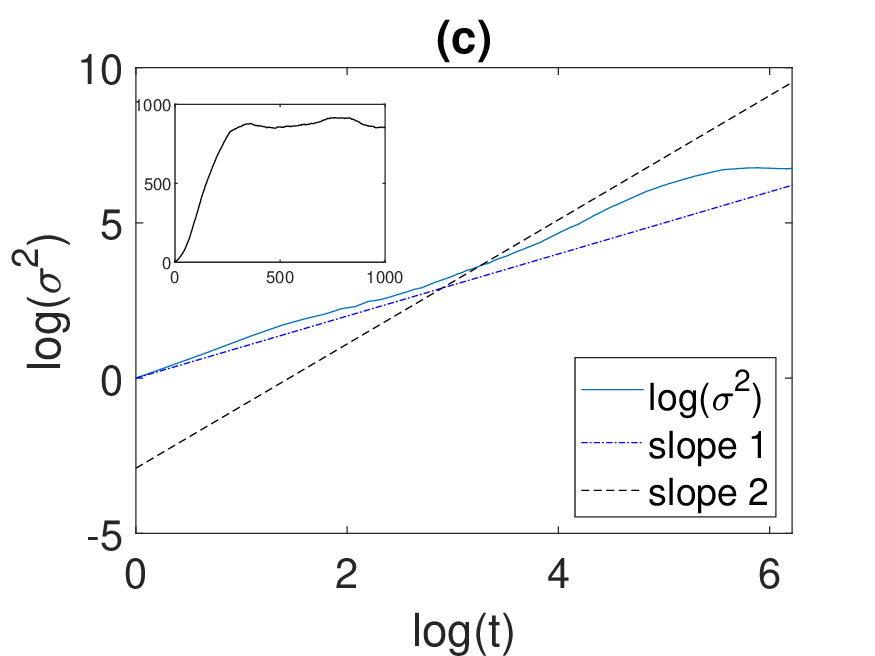} \label{fig:harper_1 c}}

   \caption{ Quantum walk with near-integrable ($g= 0.01$) and chaotic Harper ($g=0.4$) coins. (a) Probability distribution of the walker at $t=40$, shows that it approaches the well-known Gaussian distribution when the coin is chaotic, while the near-integrable case resembles a Hadamard coin. (b) Variance of the walker as a function of time, indicates the transition from ballistic to diffusive growth of the walker when the parameter $g$ changes from the near-integrable to the fully chaotic coin, here $N=101$ and $M=64$. (c) The walker variance (for the case of $g=0.4$, $N=401,$ and $M=40$) shows a transition from diffusive to ballistic growth at sufficiently long time scales. Inset shows the long-time behavior and saturation of the variance due to the finite lattice size.  }
 \label{fig:harperprob}
\end{figure*}

Turning our attention to the dynamical properties of the walker, the probability of finding the walker at site $n$ after a time $t$, $p(n,t)$, obtained by tracing out the coin is shown in Fig.~(\ref{fig:harper_1 a}) for the case when the walker starts at the lattice origin $\ket{n=0}$ and the coin state is the zero momentum state $\ket{m=0}$. For large coin dimensionality, the probability distribution during a time $t<t_c \sim M$ approaches the normal distribution when $g=0.4$ where as for $g=0.01$ it shows significant deviations. This indicates that when the coin is quantum chaotic the walker approaches the limit of a classical random walker whereas for the integrable coin the walker still behaves quantally. The above quantum to classical nature of the walker is also illustrated by studying the growth of variance, for the integrable coin the growth of the variance is ballistic with $\br n^2\kt \sim t^2$ where as for the chaotic coin the growth of variance is approximately diffusive being $\sim t$, as seen in Fig.~(\ref{fig:harper_1 b}). At long times the growth must be 
ballistic irrespective of the nature of the coin \cite{PhysRevLett.91.130602}, and indeed we see in Fig,~(\ref{fig:harper_1 c}) a transition to a ballistic growth characteristic of quantum walks such as the Hadamard walk even for the quantum chaotic case.

%
%\begin{figure*}
%\subfloat[]{\includegraphics[width =2.7in]
%{random_prob} \label{fig:prob random}}
%\subfloat[]{\includegraphics[width =2.7in]
%{random_standard} \label{fig:standard random}}
%\caption{CUE coin: Probability distribution and variance is given for the case of the random coin and it produces similar results to chaotic coin, and is to be expected.}\label{fig:proprandom}
%\end{figure*} 

\section{Random Coin and universal behavior of quantum chaotic walk}

%\subsubsection{Random Coin}
As for large $g$ we expect the dynamics to be chaotic, the coin unitary $U_C$ can be usefully replaced with a random unitary picked from the Circular unitary ensemble (CUE), which samples the unitary group uniformly. Results not presented here confirms that the probability distribution and the growth of variance of the walker position and indeed the results are identical to the one observed for the case of the chaotic regime of the kicked Harper coin using the same initial conditions.

\subsection{A path-integral formalism and odd-even effect}
For the initial state $\ket{0}_C\ket{0}_W$ the probability distribution of the walker follows from
 Eq.~\eqref{eq:RCW 2}
as (see also \citep{ErmanPazSaraceno}),
\begin{equation}
\label{eq:pnt}
\begin{aligned}
p(n,t)=&\tr[|n \kt \br n| \rho_W(t)]\\=&\frac{1}{N^2}\sum_{k,l=0}^{N-1}\bra{0}U_l^{-t}U_{k}^{t}\ket{0}\exp\left[\dfrac{2\pi i (l-k)n}{N}\right].
\end{aligned}
\end{equation}
There is an exact symmetry present in Eq.~(\ref{eq:pnt}) that for an even number of sites $N$ this restricts the walker to the even sublattice consisting of $n=0,2, \cdots$ for even times $t$ and the odd sublattice at odd times.
This is clear in the classical random walk with the start at the origin and a certainty of transiting to one of the nearest neighbours. This is inherited by the quantum walker at all times. To see this, 
use Eq.~(\ref{eq:UK}) and define $u_{1}=P_R U_c$ and $u_{-1}=P_L U_c$, and $\sigma_j =\pm 1$. Observe that 
\beq
U_k^t = \sum_{\{\sigma_j=\pm 1\}}e^{\frac{-2 \pi i k}{N}\sum_{j=1}^t \sigma_j} u_{\sigma_1}\cdots u_{\sigma_t}. 
\eeq
The double sum in Eq.~(\ref{eq:pnt}) can be carried out exactly to get 
\beq
\label{eq:pnt_path}
\begin{split}
p(n,t)&= \br 0(n,t)|0(n,t)\kt, \, \text{where}\\
|0(n,t)\kt& = \sum_{\sum_{j=1}^t\sigma_j-n =0\, \text{mod}\, N} u_{\sigma_t} \cdots u_{\sigma_1}|0\kt.
%&\sum_{\sum_{j=1}^t\sigma_j=\sum_{j=1}^t\mu_j=n} \br 0 |u_{\sigma_t} \cdots u_{\sigma_1} u^{\dagger}_{\mu_1} \cdots u^{\dagger}_{\mu_t}|0 \kt.
\end{split}
\eeq
Thus the probability  amplitude is obtained as a sum over all possible classical paths connecting the origin and site $n$ in a time $t$, and is a path integral version of the walker probability. Path integrals seem to have been applied to quantum walks 
earlier, for example \cite{konno2002quantum,konno2004path,yang2007path,joshi2018path}. Now, if there are $r$ ($+1$) and $t-r$ ($-1$) in a given binary string of length $t$, then $2r-t-n= 0 \, \text{mod}\, N$, If the number of sites, $N$, is even, the time $t$ and lattice site $n$ must have the same parity, else $p(n,t)=0$. The walk alternates between the even and odd sublattices. However, if $N$ is an odd number $2r-t-n$ could be either even or odd and hence
all sites can be occupied. This is true for the classical walker as well and we note that this is completely independent of the dynamical nature of the coin, or its symmetries. We will point to the dramatic consequence of this for the walker-coin entanglement.

\subsection{Fidelity and the walker probability}
Returning to the Eq.~(\ref{eq:pnt}) for $p(n,t)$ we note that the term $\bra{0}U_l^{-t}U_{k}^{t}\ket{0}$ for $k \neq l$ resembles the fidelity used in the Loschmidt Echo which measures the sensitivity of quantum evolution to perturbations in the Hamiltonian \citep{gorin2006dynamics}. Here the Loschmidt echo is a natural consequence of the evolutions under two different conserved momenta sectors. 
Using Eq.~(\ref{eq:UK}) and the fact that $P_R$ and $P_L$ are orthonormal projectors we get
\begin{equation}
\begin{aligned}
&U_{l}^{-1}U_k =U_{k+\Delta}^{-1}U_k\\
=& \exp(\frac{2\pi i\Delta}{N})U_C^{-1} P_R U_C+\exp(\frac{-2\pi i \Delta}{N}) U_C^{-1} P_L U_C\\
\equiv  & V_{\Delta}(0), 
\end{aligned}
\end{equation}
as the Floquet ``perturbation operator", equivalent to the modified part of the Hamiltonian in the Loschmidt echo,
and with $\Delta/N=(l-k)/N$ as the strength of the perturbation. Iterating forward to higher order in time yields,
\begin{equation}
U_l^{-t}U_{k}^{t}=V_{\Delta}(t-1) V_{\Delta} (t-2) \cdots V_{\Delta}(0),
\end{equation}
where $V_{\Delta}(t)=U^{-t}_k V_{\Delta}(0) U_k^{t}$ is the time evolved perturbation operator.
%with
%\beq
%\label{eq:Voft}
%V_{\Delta}(t)= \exp(\frac{2\pi i \Delta}{N}) P_{R}(t)+\exp(-\frac{2\pi i \Delta}{N}) P_{L}(t),
%\eeq 
%and $P_{R.L}(t)=U_k^{t}P_{R,L}U_{k}^{-t}$ are the coin projector operators evolved under the dynamics of the $U_k$.
%\red{AL: At some point we have to pay attention to the sign of $t$, forward evolution would be $U^{-t} A U^t$, maybe shift the $-t$ to $k$ in Eq. 8?}

\begin{figure*}
\subfloat{\includegraphics[width =2.8in]
{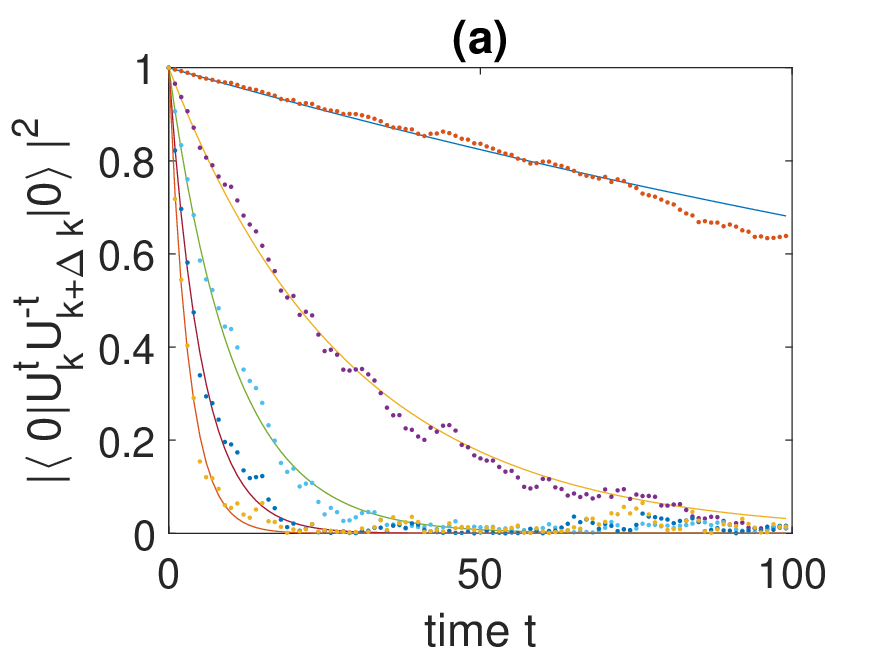} \label{comparison_random_a}}
\subfloat{\includegraphics[width =2.8in]
{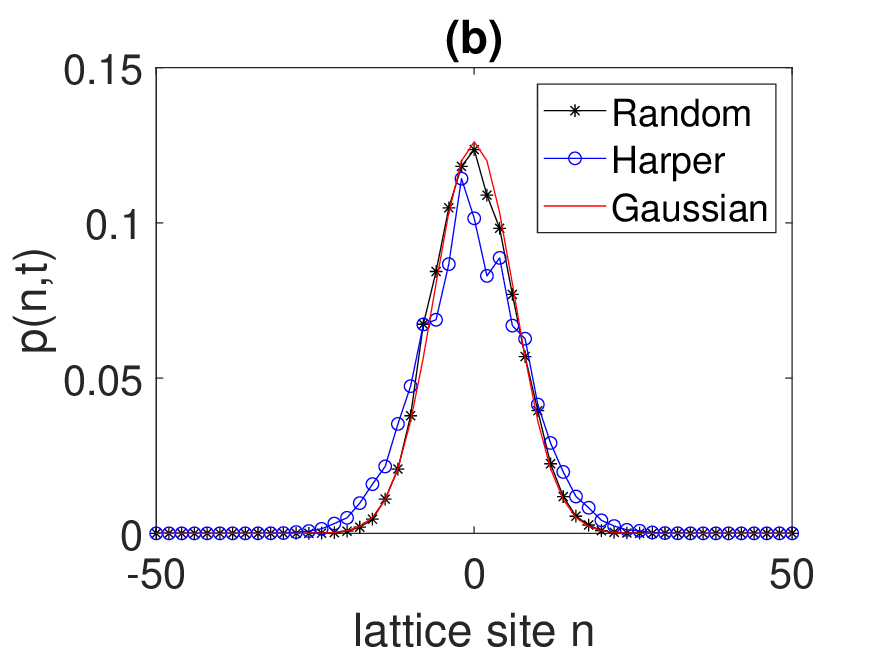}\label{comparison_random_b}}
\caption{ The figure illustrates the validity of the probability distribution obtained using the Loschmidt Echo method. (a) Fidelity decay in Eq.~\eqref{eq:UkUl} (dotted line) is compared with the exact numerical calculations (solid line) for the random coin, restricting for clarity and convenience to the first $5$ odd  values of $\Delta$. (b) Compares the actual probability distribution to the Gaussian distribution after time $t=40$, (with $N=101$, $M=256$). the probability distribution for CUE coin as well as Harper coin with $g=0.4$ is shown. }\label{fig:comparison_random}
\end{figure*}

However for the case of a random or chaotic coin matrix, there is nothing special about the initial coin state being $\ket{0}$ and hence we argue that 
 \begin{equation}
 \label{eq:xi}
\bra{0}U_l^{-t}U_{k}^{t}\ket{0}= \bra{0}V_{\Delta}(t-1)\cdots V_{\Delta}(0)\ket{0}\approx\frac{1}{M}\xi(t), 
 \end{equation}
 where $\xi(t)=\Tr[V_{\Delta}(t-1)\cdots V_{\Delta}(0)]$. Ignoring correlations between different powers of time $t$ in $U_k^{t}$, we treat them as independent CUE realizations. This allows us to average
 over say the highest power of $U_k$ that appear in $\xi(t)$ as if
 $U_k^{t-1}$  was itself a random matrix independent of any others that appear in the expression. Naturally this is an approximation and yields  
\begin{equation}
\xi(t) \approx \overline{\Tr[U_{CUE} V_{\Delta}(0) U^{\dagger}_{CUE}V_{\Delta}(t-2) \cdots V_{\Delta}(0)]}^{U_{CUE}}.
\end{equation}
We will use a basis in which $V(0)$ is diagonal and the fact from random matrix theory \cite{haake1991quantum} that 
\beq 
\overline{U_{n'm}U_{lm}^{*}}^{CUE}=\frac{1}{M}\delta_{n'l}
\eeq
to get an approximate and very simple map that is immediately solved with $\xi(0)=M$ (from Eq.~(\ref{eq:xi})):
\begin{equation}
\label{eq:cospowt}
\begin{aligned}
\xi(t) \approx &\cos\left(\frac{2 \pi \Delta}{N}\right)\xi(t-1)=M \cos^{t}\left(\frac{2 \pi \Delta}{N}\right),\\
\end{aligned}
\end{equation}
 and hence for $t \geq 0$
\begin{equation}
\label{eq:UkUl}
\bra{0}U_l^{-t}U_k^{t}\ket{0} \approx \cos^{t}\left(\frac{2 \pi \Delta}{N}\right)\approx \exp(-\frac{2\pi^2\Delta^2 t}{N^2}),
\end{equation}
where the last expression is obtained in the small perturbation limit $\Delta \ll N$. This simple derivation based on approximations as made above yields an exponential 
decay of the fidelity with a rate that is proportional to the square of the perturbation.
This is known to occur for small perturbation strengths from the general theory of the Loschmidt echo as the ``Fermi-golden-rule" regime \cite{goussev2012loschmidt,gorin2006dynamics}. 
There is an intricate set of time scales for the
Loschmidt echo, including an interesting Lyapunov regime, wherein the decay rate is the classical Lyapunov exponent. A detailed treatment taking into account the effect on the walk from the different decay regimes of the ``echo" is out of the scope of this paper.  While our derivation is independent of the literature on the echo, the Fermi-golden-rule regime is sufficient to reveal the classical random walker limit. In Fig.~(\ref{fig:comparison_random})
we demonstrate the validity of the derived exponential decay.

\subsection{Emergence of the classical binomial and normal distributions}

Using Eq.~(\ref{eq:pnt}) and the approximation to the fidelity in Eq.~(\ref{eq:cospowt}),
 the probability distribution of the walker is 
\beq
\label{eq:pntapprox1}
p(n,t)=\frac{1}{N^2}\sum_{k,l=0}^{N-1}\cos^t\left[\frac{2\pi}{N}(k-l)\right]e^{2 \pi i (k-l)n/N}
\eeq 
Expanding $\cos^t \theta$ in the binomial expansion $(e^{i \theta}+e^{-i \theta})^t/2^t$ the 
double sum can be written as the  ``absolute magnitude squared" of a single one and this results in 
\beq
\label{eq:pntapprox2}
\begin{split}
p(n,t)=&\frac{1}{2^t N^2} \sum_{r=0}^{t}\binom{t}{r}\left|\sum_{k=0}^{N-1}e^{-2\pi i k(2r -t -n)/N}\right|^2\\
&=\frac{1}{2^t} \sum_{r=0}^{t}\binom{t}{r} \delta[(2r-t-n)\, \text{mod}\, N,0].
\end{split}
\eeq 
If $t < N$, only the terms $r=(t+n)/2$ and $r=(t+n-N)/2$ will contribute for $-t \leq n\, \text{mod}\, N \leq t$, provided they are integers. For convenience identify the walker lattice index $n$ with $n-N<0$ when $n>N/2$, so as to center the starting site $n=0$. The simple classical symmetric random walk on the infinite one-dimensional lattice then results from the above expression explicitly as,
\beq
\label{eq:pntapprox3}
p(n,t)=\frac{1}{2^{t+1}}[1+(-1)^{n+t}]\binom{t}{\frac{t+n}{2}}.
\eeq
with $-t \leq n \leq t$.
Thus we recover the well-known binomial distribution of the position of the classical 
walker \cite{ChandraRMP1943}, with the feature that the site $n$ and time $t$ have the same parity intact.
Indeed $\cos^t(\theta)$ is the characteristic function of the classical random walk and appears
as the fidelity approximation of the quantum walk derived above. It is now standard to recover the normal or Gaussian approximation
from the binomial for $t \gg n$, and in practice $t$ can be as small as $10$ \cite{ChandraRMP1943}:
\beq
\label{eq:normal}
p(n,t)=\frac{1}{2}[1+(-1)^{n+t}]\, \sqrt{\frac{2}{\pi t}}e^{-n^2/2t}.
\eeq
The comparison  of the probability distribution obtained using the above expression with the numerical calculation is given in the Fig.(\ref{fig:comparison_random}).

In order to get the above expression for probability distribution, the unitary matrix in the coin space was taken as a typical member of the circular unitary ensemble, CUE, which is true for the quantization of chaotic maps in general and hence the above fact implies that the results are quite general in nature and for any quantum chaotic walk.

%\textcolor{blue}{} 
 
\subsection{Time scales} 
\begin{figure}[]
%\subfloat[$M\gg N$]{\includegraphics[width =1.8in]{M_largerthan_N}}
%\subfloat[$M\gg N$]{\includegraphics[width =1.8in]{entanglement_33}}
%\subfloat[$M<<N, M<30$]{\includegraphics[width =2.7in]{N_larger_than_M}}
\includegraphics[scale=.5]{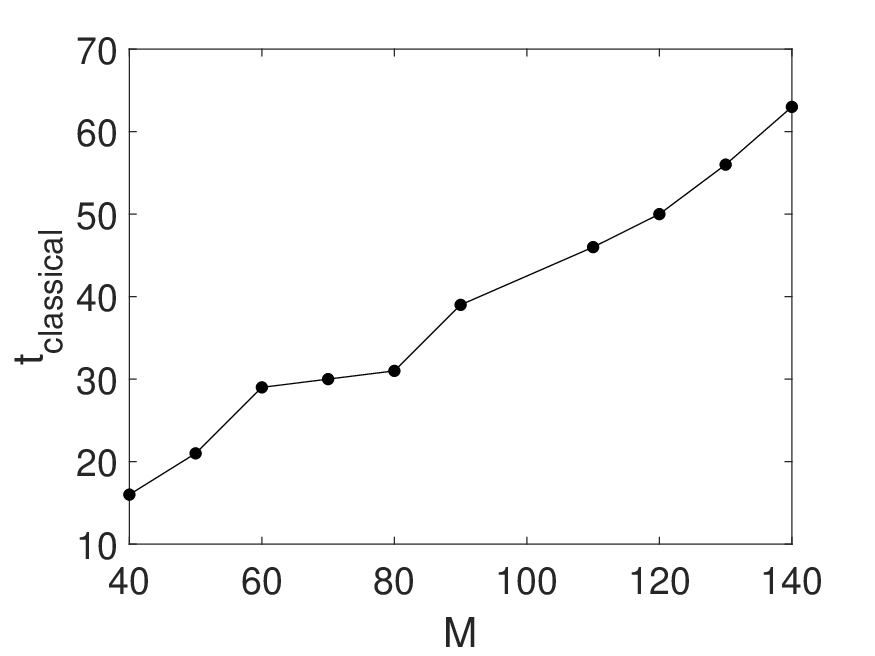}
%\subfloat[Entanglement]{\includegraphics[width =1.8in]{compare_variance_entanglement}}
\caption{The diffusive time $t_{D}$ during which the walker diffuses classically, \emph{vs} $M$ for $N= M+1$. $t_{D}$ is obtained numerically as the time at which the variance deviates from its diffusive growth as for example in Fig.(\ref{fig:harperprob}c). } \label{fig:time_scale0}
\end{figure}

Thus we see that there is an extended classical diffusive behavior of the walker which goes far beyond the classical-quantum correspondence time or the Ehrenfest time of the $t_{EF}$  of the coin alone. This time scale is well-known to depend on whether the system, in this case the coin dynamics, is integrable or not, being much shorter for nonintegrable chaotic ones \cite{berry1979evolution,zurek2001sub} and scales as $t_{EF}=\ln M/\lambda$, where $\lambda$ is the Lyapunov exponent of the classical limit of the coin.Thus while quantum effects set into the coin subsystem rather early, the diffusive nature of the walker 
which is also ``classical" lasts much longer. However, the walker is strongly coupled to the coin via the controlled operations, and its classical
behaviour is a result of this and lasts till a ``diffusive" time scale. This diffusive time scale, $t_D$ was earlier studied in \citep{ErmanPazSaraceno}, who found it to scale as $\mathcal{O}(M)$ and Fig.~(\ref{fig:time_scale0}) shows that for $M \leq N$ the $t_{D}$ is of $\mathcal{O}(M/2)$ and hence is independent of the walker dimension and $\gg t_{EF}$. From our treatment of how the normal or binomial distribution arises, it is clearly governed by the timescales at which the fidelity saturates from its exponential decay in \eqref{eq:UkUl}. It maybe noted that in the case of a non-chaotic (with a two dimensional quantum coin) discrete quantum walker whose initial states were coherent states, it was observed that the Ehrenfest time scales as $\sqrt{N}$ where $N$ is the dimension of the lattice \cite{omanakuttan2018quantum}.
 
\begin{figure}[]
\includegraphics[scale=.5]{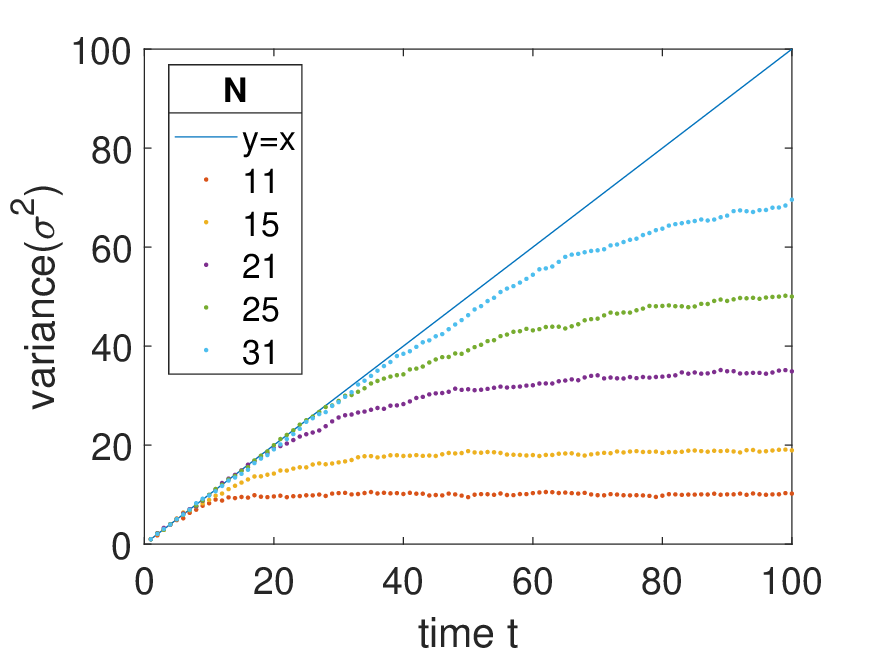}
   \caption{ Illustrates the saturation of variance for the coin dominated case ($M\gg N$). The linear growth of variance in the initial phase of the walker paves way to saturation rather than ballistic growth and the case is illustrated for different values of $N$, when $M=100$.}\label{fig:time_scale11}
\end{figure}

But there is an interesting class of walkers, who do not achieve a quantum phase of growth. This is when $M\gg N$, that is the lattice is much smaller than the coin dimensionality. In this case by the time the diffusive behaviour has ended, the compactness of the walker space becomes important. Thus this is 
the case of finite phase space and the cyclic nature of the graph dominates. In these cases we observe a classical diffusive growth giving way to saturation after about $t=t_D$, as shown in Fig.~(\ref{fig:time_scale11}).

\section{Coin-Walker Entanglement}

\begin{figure*}
\subfloat{\includegraphics[width =2.3in]
{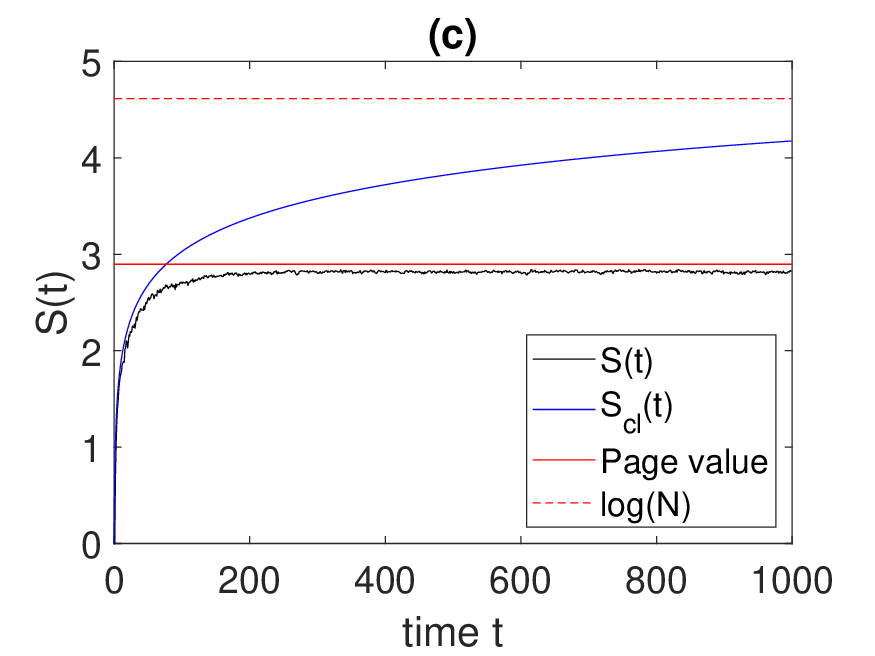}
\label{fig:entropya}}
\subfloat{\includegraphics[width =2.3in]
{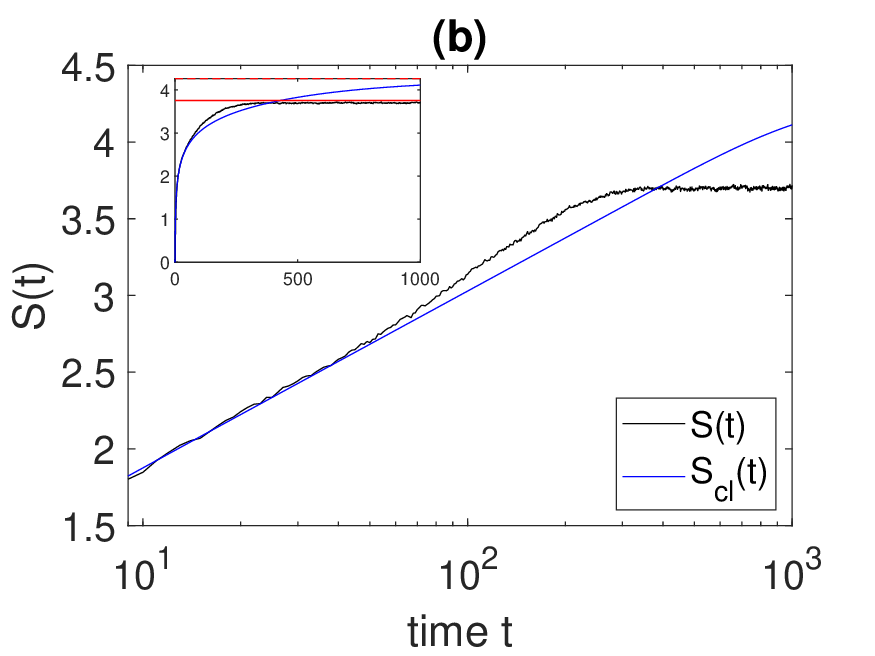}\label{fig:entropyb}}
\subfloat{\includegraphics[width =2.3in]
{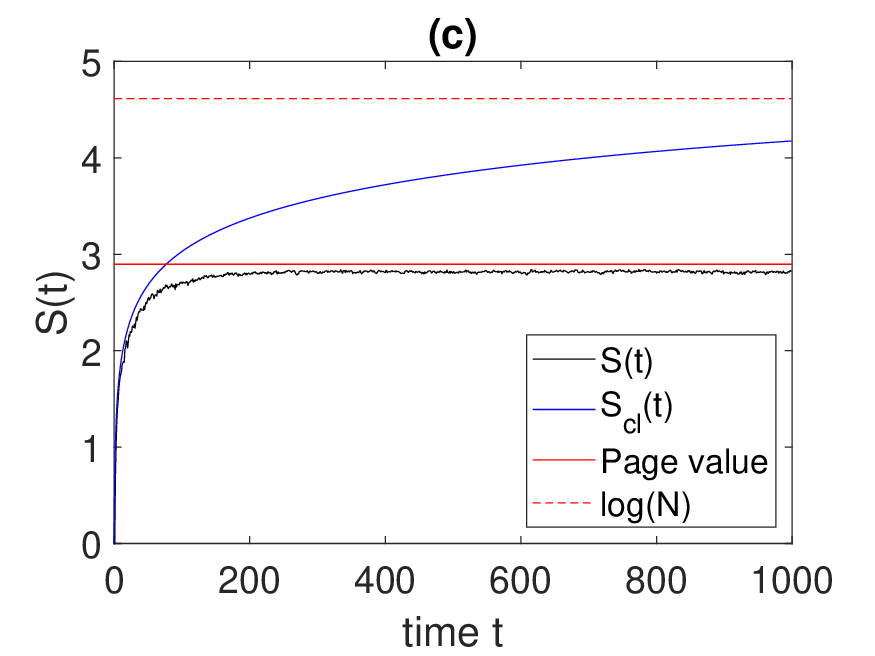}\label{fig:entropyc}}
\caption{Entanglement, as measured by the von Neumann entropy, as a function of time for different cases with random coin. The saturation is compared to the Page value, horizontal lines, which is the statistical average over the Haar measure of pure states in $MN$ dimensions.(a) coin dominated case: $M\gg N$ with $M=100$ and $N=21$, (b) $M=70$ and $N=71$ and (c) walker dominated case: $M\ll N$ with $M=20$ and $N=101$. The central figure is in log scale to illustrate the closeness of the actual curve to the one obtained from the classical probability distribution as given by Eq.\eqref{eq:classical_entropy} and the inset is in actual scale with same axis. In cases (a) and (b)the entropy saturates to the Page value indicating that the final state is close to a random state.}
\label{fig:entropy_fig_combined}
\end{figure*}

In the context of quantum chaotic walks, linear entropy which is an entanglement monotone was previously studied in \cite{ErmanPazSaraceno}, and
unless otherwise specified we use the von Neumann entropy of the reduced density matrices as the measure of entanglement:
\beq
\label{eq:Sent}
S(t)=S(\rho_{W}(t))=S(\rho_C(t))=-
\tr[ \rho_W(t) \ln \rho_{W}(t)],
\eeq
where $\rho_W(t)$ and $\rho_C(t)$ are given explicitly in Eqs.~(\ref{eq:RCW}) and ~(\ref{eq:RCW 1}).
Entanglement of two (bipartite) coupled chaotic systems has been studied for a while, a sample being \cite{MillerSarkar1999, PhysRevE.64.036207,bandyopadhyay2002testing,Chaudhury2009,Neill16} and it is known that for sufficient chaos and interactions, the 
entanglement can reach that of random states, which is nearly maximal \cite{bandyopadhyay2002testing,BandyoLak2004,Fujisaki03}. The quantum walk studied in this paper presents an
intriguing variation, wherein one subsystem, namely the coin's is potentially chaotic, but the walker dynamics is simply a shift or free particle.
One may consider the interaction to be determined by the probability of the transitions to the left or right. For example in Eq.~(\ref{eq:qwU})
if $P_R=\mathds{1}_M$ and $P_L=0$, the walker always goes to the right and the walker-coin system is evidently decoupled. This is also in the case when 
$P_L=\mathds{1}_M$ and $P_L=0$, and the case we consider is in this sense of maximum interaction with both of them being equally likely.

If the bipartite systems are both fully chaotic, the time development of the entanglement of uncoupled eigenstates has been recently studied in detail \cite{pulikkottil2020entanglement}. For general initial product states earlier results include a rapid saturation to the random state value \cite{Fujisaki03,BandyoLak2004}, including a linear regime for very weakly interacting cases \cite{MillerSarkar1999}. Generically a linear
entropy increase is expected even for integrable systems such as the transverse field Ising model \cite{Calabrese2007}.
In the case of the quantum walk, it is interesting that the entanglement develops much more slowly, probably originating in the mixed nature of the dynamics of the walker and the coin, and the conservation of the walker momentum.

As a first estimate of the entanglement, we may use the classical entropy. Indeed, the 
expression for $S(t)$ maybe obtained in the ``classical regime" when we use the approximation in Eq.~\eqref{eq:UkUl} to obtain
the walker state of Eq.~(\ref{eq:RCW 1}). It is not hard to see with a similar calculation as for the diagonal elements
of this density matrix that obtained $p(n,t)$ in Eq.~(\ref{eq:pntapprox2}) that $\br n |\rho_W(t)|m \kt=0$ if $n \neq m$.
That is, under the approximation wherein we derive the classical random walk, the walker density matrix is purely diagonal. Thus the eigenvalues
of the reduced density matrix are approximately simply the probability of site occupancies, $p(n,t)$. This in turn implies that the entanglement entropy is close to the classical Shannon entropy
\beq
\label{eq:Clentropy}
S(t)\approx S_{\text{cl}}(t)=-\sum_{n=0}^{N-1} p(n,t) \log p(n,t).
\eeq
If we use the binomial distribution in Eq.~(\ref{eq:pntapprox3}) for $p(n,t)$, valid for $t<N$, then 
using Stirling's approximation valid for $t,N\gg 1$, we get 
\begin{equation}
S_{\text{cl}}(t) \approx \frac{1}{2} \log(\frac{\pi e t}{2}).
\label{eq:classical_entropy}
\end{equation}
Hence the growth of the entanglement, even with a fully chaotic walker, is only logarithmic as opposed
to linear and occurs over a long time scale. Significantly, other physical situations have given rise to logarithmic 
entanglement growth, such as in the many-body localized phases of spin chains \cite{Bardarson2012,Znidaric2008,serbyn2013universal},
in chaotic quantum coupled kicked rotors \cite{Park2003}, decohering kicked rotors \cite{Nag2001} (in both of these
the growth is exactly as in the case of quantum walks a classical growth $\sim \log(\sqrt{t})$) and in 2D disordered
free fermion systems \cite{Zhao2019}.

The simplest case, when the classical entropy is essentially the entanglement is the coin dominated 
one when $M \gg N$. In all of this discussion we only consider chaotic or random coins. 
The entanglement saturates shortly after the classical phase and the logarithmic growth happens.
The classical entropy goes on to the maximal value of $\log N$ when $N$ is an odd integer so that the bipartite lattice
symmetry is broken. The quantum entanglement closely follows the classical curve but saturates at a slightly lower
value. This is due to the formation of a random state in the product $MN$ dimensional space, which is as if it were 
picked from the uniform Haar measure. Thus the coin-walker system eventually thermalizes to a combined random state, despite the walker's
simple nearest neighbor hopping dynamics and the conservation of lattice momentum: the coin's chaoticity is sufficiently dominant.

Fig.~(\ref{fig:entropya}) shows the quantum entanglement along with the classical entropy for a coin dominated. Their closeness throughout
all phases is remarkable. The growth phase of $\sim \log(\sqrt{t})$ gives way to saturation when the walker folds over
the lattice and thermalizes. While the classical entropy is very close to $\log N$, the quantum entropy approaches the so-called Page value \cite{page1993average}. The exact value of the average entanglement of random $N_1 N_2$ dimensional pure states on $\mathcal{H}^{N_1}\otimes \mathcal{H}^{N_2}$ with $N_1\leq N_2$ was conjectured by Page in \cite{page1993average} and later proved by others, for example see \cite{sen1996average}. It is approximately given for $N_2\geq N_1 \gg 1$ by, 
 \begin{equation}
 \label{eq: Page value}
 S_{\text{Page}} \approx \ln N_1-\frac{N_1}{2N_2}.
 \end{equation}
Fig.~(\ref{fig:entropyb}) is for the case when the coin and walker spaces are nearly isomorphic, $M\approx N$, and we see that the 
Page value is still reached, but also interestingly the quantum entanglement at late times can be even more than the classical entropy.
The walker dominated case of $N \gg M$ is shown in Fig.~(\ref{fig:entropyc}) and here the classical entropy is much larger as it approaches
$\log N$, while the entanglement can be at most $\log M$. The entanglement does not also reach the corresponding Page values and hence
the states are not typical even after saturation.
 
\begin{figure*}
\subfloat{\includegraphics[width =2in]{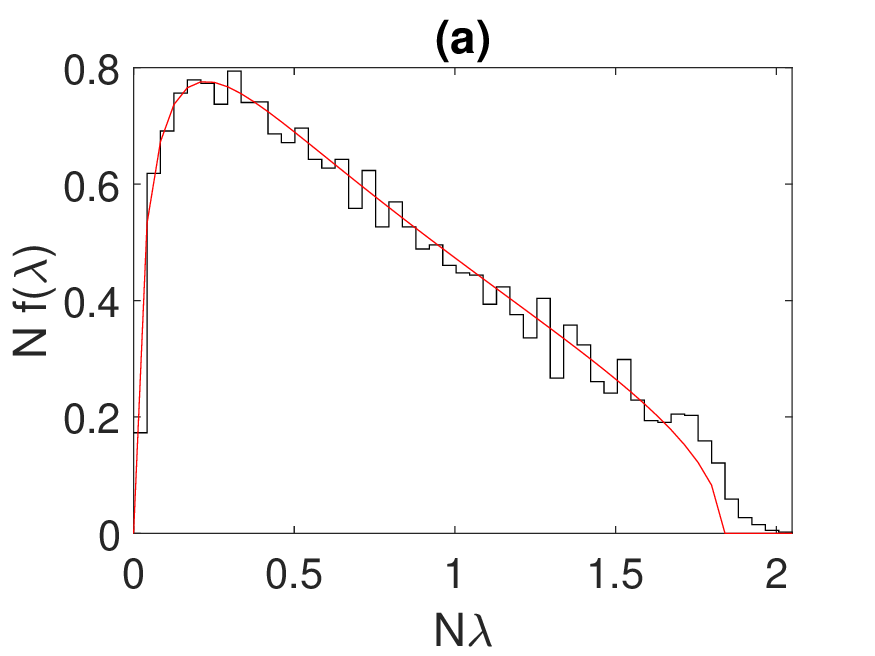}
\label{fig:MPa}}
\subfloat{\includegraphics[width =2in]{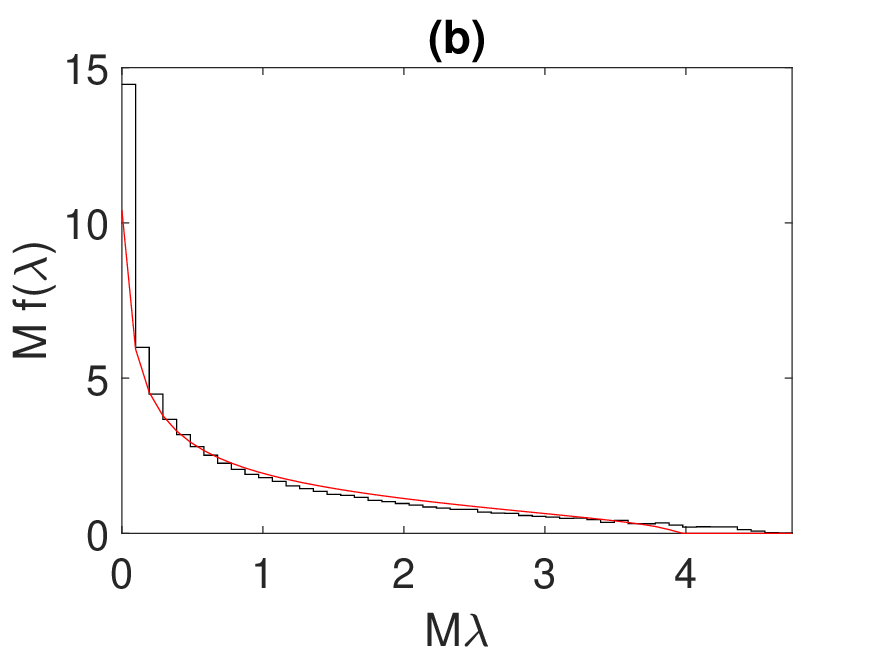}
\label{fig:MPb}}
\subfloat{\includegraphics[width =2in]{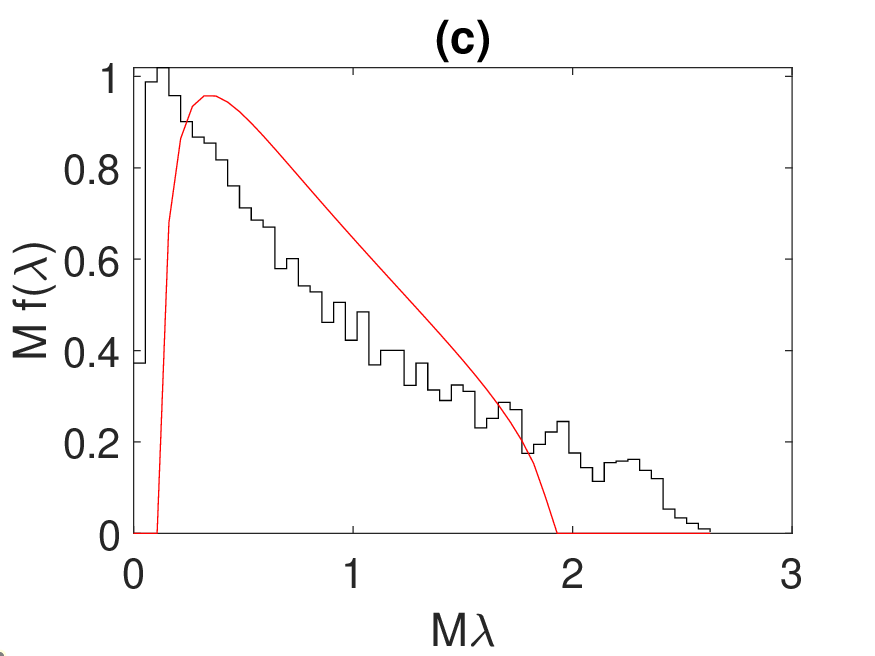}
\label{fig:MPc}}
\caption{Distribution of the eigenvalues of the walker or coin reduced density matrices from a time $t=700$ to $t=1700$ of the evolution. Solid  curves corresponds to the Marcenko-Pastur law in Eq.\eqref{eq:MP distribution}. (a) coin dominated case: $M\gg N$ with $M=100$ and $N=21$, (b) $M=70$ and $N=71$ and (c) walker dominated case: $M\ll N$ with $M=20$ and $N=101$. }
\label{fig:MP}
\end{figure*}

To ascertain if random states are reached asymptotically, we also find the 
distribution of the spectra of the reduced density matrices. If the entangled state is a random one on the product space as above, 
the distribution of these eigenvalues will follow the 
Marchenko-Pastur (MP) law \cite{Marcenko67, BZBook} given as 
\begin{equation}
\begin{aligned}
f(\lambda)=&\frac{N_1Q}{2\pi }\frac{\sqrt{(\lambda-\lambda_{\text{min}})(\lambda_{\text{max}}-\lambda)}}{\lambda};\\
\lambda_{\text{min}}^{\text{max}}=&\frac{1}{N_1}\left(1+\frac{1}{Q}\pm\frac{2}{\sqrt{{Q}}}\right)
\end{aligned}
\label{eq:MP distribution}
\end{equation}
where $Q=N_2/N_1$. The MP law has been found to hold from bipartite quantum chaotic systems, for example \cite{bandyopadhyay2002testing,Kubotani2008}, to many-body systems such as in \cite{Mishra2014,Parisi17}. Figure~(\ref{fig:MP})
shows three cases as discussed above and we see that the MP distribution is an excellent approximation for the coin dominated 
case, but not for the walker dominated one. Indeed, note that the interchange of coin and walker dimensionalities completely changes
the distribution. It will be interesting to study how these more general distributions arise, and the role of conservation 
laws.

\begin{figure*}
\subfloat{\includegraphics[width =2.5in]{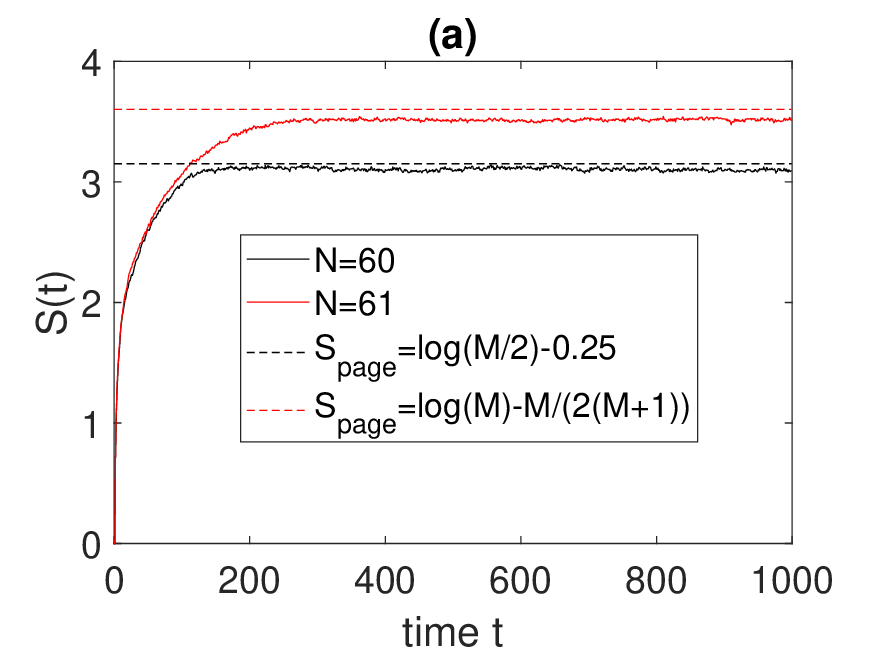} \label{fig:odd-even asymmetry a}}
\subfloat{\includegraphics[width =2.5in]{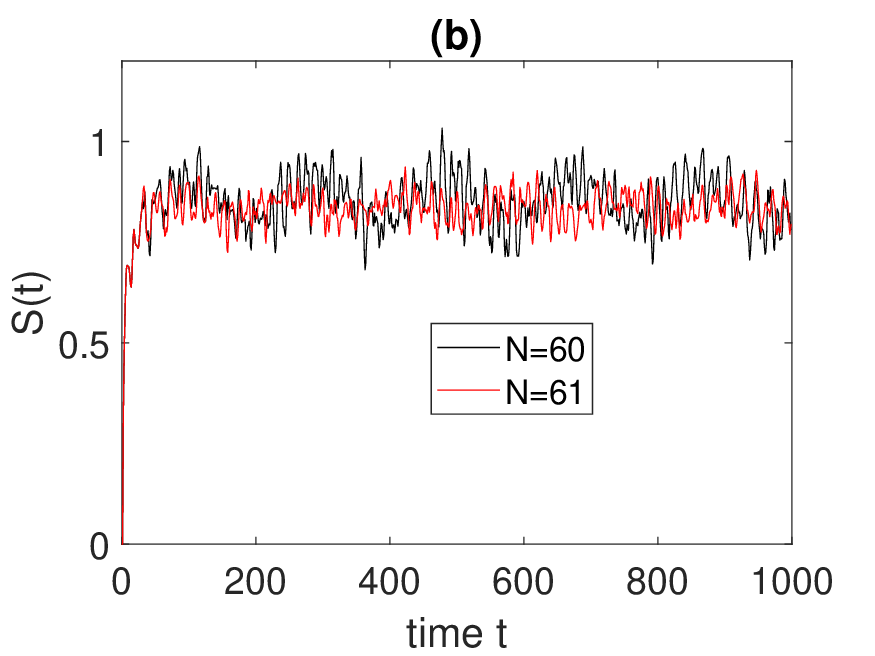}  \label{fig:odd-even asymmetry b}}
\caption{Entanglement, as measured by the von Neumann entropy, as function of time. The saturation value of the entanglement indicates the difference between odd and even values of $N$ for $M=60$. (a) Random coin, the saturation value of entanglement is different for $N$ odd and even. (b) Harper coin with $g=0.001$, which is well below the chaotic critical point of $g=0.05$ and the saturation value of $N$ odd and $N$ even are approximately same. Thus the value of even or odd $N$ is significant for the chaotic/random coin however, for the integrable case there is not much of a distinction. }
\label{fig:odd-even asymmetry}
\end{figure*}

The walker dimension we have used in these calculations have all been odd integers. This is due to the even-odd effect, originating from
the bipartite lattice symmetry being obeyed by the walker, both classical and quantum as pointed out earlier.
To recall, from Eq.\eqref{eq:pnt_path}, the probability distribution of the walker is obtained as a sum of all classical path and hence in the long time limit, similar to the classical random walk for $N$ odd all the sites can be occupied whereas for $N$ even only half the sites can be.  Also, for the case of a chaotic/random coin all the degrees of freedom are accessed and hence the entanglement only depends on the degrees of freedom of the walker. Hence, for $N$ even and $M=N$ the average value of the von-Neumann entropy is given by Eq.\eqref{eq: Page value} with $N_1=M/2$ and $N_2=M$, $S_{\text{Page}}=\log(M/2)-.25$. However for the case of $N$ odd (and $N=M+1$) the entanglement nearly achieves the Page value with $N_1=M$ and $N_2=M+1$, $S_{\text{Page}}=\log(M)-M/(2 (M+1))$. Thus the odd-even effect has dramatic consequences for the saturation entropy and and is shown in Fig.~(\ref{fig:odd-even asymmetry a}). While this may not be that surprising, what is interesting however is that if the coin is the quantization of the integrable classical dynamics there is no major distinction between the odd and even lattice sites,
as shown in Fig.~(\ref{fig:odd-even asymmetry b}). This in turn is due to the fact that not all the degrees of freedom of the coin are accessible for the integrable case and random states are not obtained, although there does seem to be an equilibrium value of the entropy with large fluctuations. 

\section{Summary and discussions}

This work has studied coined quantum walks on a simple one-dimensional lattice with periodic boundary conditions.
The coin was taken as a quantum chaotic system, but with a parameter than can also allow for integrable and 
intermediate dynamics. We have pointed out that a Loschmidt echo like fidelity construct is central to this study. By deriving 
this approximately using random matrix theory, we have shown how the classical random walk's well-known binomial distribution
and the normal distribution arise when the coin is chaotic. Interestingly the fidelity is precisely the classical 
walker's characteristic function. While quantum walks have been studied now for nearly 20 years, to our knowledge 
this is not explicit, although the normal diffusion has been derived.
It may also be interesting to look at the Loschmidt echo aspect for better approximations that go beyond 
the classical one, as the literature on the echo is extensive. We have thus made connections with this well-studied 
area of quantum chaos and quantum walks. We emphasize that that we take finite walker spaces and allow for the coin 
dimensionality to vary, and the classical limit is its limit to $\infty$, when indeed the classical coin achieves
randomness via deterministic chaos.

We have also pointed out a path-integral formalism of the quantum walk that can be used to define
walks on arbitrary graphs and that helps in seeing the role of the bipartite lattice symmetry, with odd and even
dimensional lattices behaving very differently if the coin is sufficiently chaotic. Entanglement between
the coin and walker was also studied and showed some intriguing features. Starting from an logarithmic growth
as dictated by the diffusive classical limit, it saturates to a steady state value. This steady state value
as well as the nature of the states accessed in this regime depends crucially on the relative sizes
of the coin and walker spaces. It was shown that for coin dominated cases, the chaos is sufficient to
drive the whole system to a typical random state, such that the reduced density matrix eigenvalues 
have the universal Marchenko-Pastur distribution and the entanglement reaches the Page value. However in the
walker dominated case, this is not true, and even for large coin dimensions ($M \gg 2$), the Page value is 
not reached. In this regime the conservation of lattice momentum due to translation symmetry seems to 
play a crucial role and it will be both interesting to see if one could derive these as well as
to see how generic these distributions are. This may be related to ways in which states evolve in 
general complex systems, a recent survey is found in \cite{volya2020time}.
It is well-known that many-body systems constrained 
with symmetries such as particle number, magnetization, or even just energy, can show 
deviations from such universality.  

Discrete time coined walk has been implemented in the NISQ devices \cite{georgopoulos2019comparison,chen2019hybrid} and shown to exhibit quantum nature of the walker. In this context it will be interesting to see whether one can implement a quantum chaotic walk and show the transition in the behavior of the walker depending on the integrability of the coin. Thus providing a simple and elegent test bed for the effects of quantum chaos of subsystem dynamics and providing a platform to further explore the role of quantum chaos in quantum computation.
Also, It will also be interesting to understand whether one  can extend quantum chaotic walk to the case of continuous time quantum walk and see whether one could see similar behavior.

Recently a possible connection is established between the Loschmidt echo and out of time order correlators \cite{yan2020information}. In Eq.~\eqref{eq:UkUl} we saw the natural emergence of the Loshmidt  echo for the quantum walk and expect an exponential growth of OTOCs in the coin-dominated chaotic cases, in contrast to the quadratic growth observed for the Hadamard walk \cite{omanakuttan2019out}. One of the open question is about the exact time at which the walker deviates from classical diffusive growth, studies of participation ratio and variance suggests that the time is of the order of the  coin dimension, however the transition is difficult to pin down, and it might useful to look into
newer measures of non-classicality. Another interesting question is to study the influence of decoherence, as well
as the breaking of translational symmetry. We have largely ignored the integrable and near-integrable coin cases, which could lead to 
very different types of quantum transport.

\begin{acknowledgments}
SO would like to thank Gopikrishnan Muraleedharan (Los Alamos National Laboratory) for his help during various stages of the progress of the work. Additionally, SO thanks the members of the Center for Quantum Information and Control (CQuIC) for their support and for discussions throughout this work.
AL was partially supported by the Department of Science and Technology, Govt. of India, under grant number DST/ICPS/QuST/Theme-3/2019/Q69.
\end{acknowledgments}

\bibliography{higherdimension}

\begin{thebibliography}{60}%
\makeatletter
\providecommand \@ifxundefined [1]{%
 \@ifx{#1\undefined}
}%
\providecommand \@ifnum [1]{%
 \ifnum #1\expandafter \@firstoftwo
 \else \expandafter \@secondoftwo
 \fi
}%
\providecommand \@ifx [1]{%
 \ifx #1\expandafter \@firstoftwo
 \else \expandafter \@secondoftwo
 \fi
}%
\providecommand \natexlab [1]{#1}%
\providecommand \enquote  [1]{``#1''}%
\providecommand \bibnamefont  [1]{#1}%
\providecommand \bibfnamefont [1]{#1}%
\providecommand \citenamefont [1]{#1}%
\providecommand \href@noop [0]{\@secondoftwo}%
\providecommand \href [0]{\begingroup \@sanitize@url \@href}%
\providecommand \@href[1]{\@@startlink{#1}\@@href}%
\providecommand \@@href[1]{\endgroup#1\@@endlink}%
\providecommand \@sanitize@url [0]{\catcode `\\12\catcode `\$12\catcode
  `\&12\catcode `\#12\catcode `\^12\catcode `\_12\catcode `\%12\relax}%
\providecommand \@@startlink[1]{}%
\providecommand \@@endlink[0]{}%
\providecommand \url  [0]{\begingroup\@sanitize@url \@url }%
\providecommand \@url [1]{\endgroup\@href {#1}{\urlprefix }}%
\providecommand \urlprefix  [0]{URL }%
\providecommand \Eprint [0]{\href }%
\providecommand \doibase [0]{http://dx.doi.org/}%
\providecommand \selectlanguage [0]{\@gobble}%
\providecommand \bibinfo  [0]{\@secondoftwo}%
\providecommand \bibfield  [0]{\@secondoftwo}%
\providecommand \translation [1]{[#1]}%
\providecommand \BibitemOpen [0]{}%
\providecommand \bibitemStop [0]{}%
\providecommand \bibitemNoStop [0]{.\EOS\space}%
\providecommand \EOS [0]{\spacefactor3000\relax}%
\providecommand \BibitemShut  [1]{\csname bibitem#1\endcsname}%
\let\auto@bib@innerbib\@empty
%</preamble>
\bibitem [{\citenamefont {Preskill}(2018)}]{preskill2018quantum}%
  \BibitemOpen
  \bibfield  {author} {\bibinfo {author} {\bibfnamefont {J.}~\bibnamefont
  {Preskill}},\ }\href@noop {} {\bibfield  {journal} {\bibinfo  {journal}
  {Quantum}\ }\textbf {\bibinfo {volume} {2}},\ \bibinfo {pages} {79} (\bibinfo
  {year} {2018})}\BibitemShut {NoStop}%
\bibitem [{\citenamefont {Arute}\ \emph {et~al.}(2019)\citenamefont {Arute},
  \citenamefont {Arya}, \citenamefont {Babbush}, \citenamefont {Bacon},
  \citenamefont {Bardin}, \citenamefont {Barends}, \citenamefont {Biswas},
  \citenamefont {Boixo}, \citenamefont {Brandao}, \citenamefont {Buell} \emph
  {et~al.}}]{arute2019quantum}%
  \BibitemOpen
  \bibfield  {author} {\bibinfo {author} {\bibfnamefont {F.}~\bibnamefont
  {Arute}}, \bibinfo {author} {\bibfnamefont {K.}~\bibnamefont {Arya}},
  \bibinfo {author} {\bibfnamefont {R.}~\bibnamefont {Babbush}}, \bibinfo
  {author} {\bibfnamefont {D.}~\bibnamefont {Bacon}}, \bibinfo {author}
  {\bibfnamefont {J.~C.}\ \bibnamefont {Bardin}}, \bibinfo {author}
  {\bibfnamefont {R.}~\bibnamefont {Barends}}, \bibinfo {author} {\bibfnamefont
  {R.}~\bibnamefont {Biswas}}, \bibinfo {author} {\bibfnamefont
  {S.}~\bibnamefont {Boixo}}, \bibinfo {author} {\bibfnamefont {F.~G.}\
  \bibnamefont {Brandao}}, \bibinfo {author} {\bibfnamefont {D.~A.}\
  \bibnamefont {Buell}},  \emph {et~al.},\ }\href@noop {} {\bibfield  {journal}
  {\bibinfo  {journal} {Nature}\ }\textbf {\bibinfo {volume} {574}},\ \bibinfo
  {pages} {505} (\bibinfo {year} {2019})}\BibitemShut {NoStop}%
\bibitem [{\citenamefont {Georgeot}\ and\ \citenamefont
  {Shepelyansky}(2000)}]{PhysRevE.62.3504}%
  \BibitemOpen
  \bibfield  {author} {\bibinfo {author} {\bibfnamefont {B.}~\bibnamefont
  {Georgeot}}\ and\ \bibinfo {author} {\bibfnamefont {D.~L.}\ \bibnamefont
  {Shepelyansky}},\ }\href {\doibase 10.1103/PhysRevE.62.3504} {\bibfield
  {journal} {\bibinfo  {journal} {Phys. Rev. E}\ }\textbf {\bibinfo {volume}
  {62}},\ \bibinfo {pages} {3504} (\bibinfo {year} {2000})}\BibitemShut
  {NoStop}%
\bibitem [{\citenamefont {Frahm}\ \emph {et~al.}(2004)\citenamefont {Frahm},
  \citenamefont {Fleckinger},\ and\ \citenamefont
  {Shepelyansky}}]{frahm2004quantum}%
  \BibitemOpen
  \bibfield  {author} {\bibinfo {author} {\bibfnamefont {K.~M.}\ \bibnamefont
  {Frahm}}, \bibinfo {author} {\bibfnamefont {R.}~\bibnamefont {Fleckinger}}, \
  and\ \bibinfo {author} {\bibfnamefont {D.~L.}\ \bibnamefont {Shepelyansky}},\
  }\href@noop {} {\bibfield  {journal} {\bibinfo  {journal} {The European
  Physical Journal D-Atomic, Molecular, Optical and Plasma Physics}\ }\textbf
  {\bibinfo {volume} {29}},\ \bibinfo {pages} {139} (\bibinfo {year}
  {2004})}\BibitemShut {NoStop}%
\bibitem [{\citenamefont {Childs}(2009)}]{childs2009universal}%
  \BibitemOpen
  \bibfield  {author} {\bibinfo {author} {\bibfnamefont {A.~M.}\ \bibnamefont
  {Childs}},\ }\href@noop {} {\bibfield  {journal} {\bibinfo  {journal} {Phys.
  Rev. Lett.}\ }\textbf {\bibinfo {volume} {102}},\ \bibinfo {pages} {180501}
  (\bibinfo {year} {2009})}\BibitemShut {NoStop}%
\bibitem [{\citenamefont {Lovett}\ \emph {et~al.}(2010)\citenamefont {Lovett},
  \citenamefont {Cooper}, \citenamefont {Everitt}, \citenamefont {Trevers},\
  and\ \citenamefont {Kendon}}]{lovett2010universal}%
  \BibitemOpen
  \bibfield  {author} {\bibinfo {author} {\bibfnamefont {N.~B.}\ \bibnamefont
  {Lovett}}, \bibinfo {author} {\bibfnamefont {S.}~\bibnamefont {Cooper}},
  \bibinfo {author} {\bibfnamefont {M.}~\bibnamefont {Everitt}}, \bibinfo
  {author} {\bibfnamefont {M.}~\bibnamefont {Trevers}}, \ and\ \bibinfo
  {author} {\bibfnamefont {V.}~\bibnamefont {Kendon}},\ }\href@noop {}
  {\bibfield  {journal} {\bibinfo  {journal} {Phys. Rev. A}\ }\textbf {\bibinfo
  {volume} {81}},\ \bibinfo {pages} {042330} (\bibinfo {year}
  {2010})}\BibitemShut {NoStop}%
\bibitem [{\citenamefont {Childs}\ and\ \citenamefont
  {Goldstone}(2004)}]{childs2004spatial}%
  \BibitemOpen
  \bibfield  {author} {\bibinfo {author} {\bibfnamefont {A.~M.}\ \bibnamefont
  {Childs}}\ and\ \bibinfo {author} {\bibfnamefont {J.}~\bibnamefont
  {Goldstone}},\ }\href@noop {} {\bibfield  {journal} {\bibinfo  {journal}
  {Phys. Rev. A}\ }\textbf {\bibinfo {volume} {70}},\ \bibinfo {pages} {022314}
  (\bibinfo {year} {2004})}\BibitemShut {NoStop}%
\bibitem [{\citenamefont {Aaronson}\ and\ \citenamefont
  {Ambainis}(2003)}]{aaronson2003quantum}%
  \BibitemOpen
  \bibfield  {author} {\bibinfo {author} {\bibfnamefont {S.}~\bibnamefont
  {Aaronson}}\ and\ \bibinfo {author} {\bibfnamefont {A.}~\bibnamefont
  {Ambainis}},\ }in\ \href@noop {} {\emph {\bibinfo {booktitle} {Foundations of
  Computer Science, 2003. Proceedings. 44th Annual IEEE Symposium on}}}\
  (\bibinfo {organization} {IEEE},\ \bibinfo {year} {2003})\ pp.\ \bibinfo
  {pages} {200--209}\BibitemShut {NoStop}%
\bibitem [{\citenamefont {Aharonov}\ \emph {et~al.}(1993)\citenamefont
  {Aharonov}, \citenamefont {Davidovich},\ and\ \citenamefont
  {Zagury}}]{aharonov1993quantum}%
  \BibitemOpen
  \bibfield  {author} {\bibinfo {author} {\bibfnamefont {Y.}~\bibnamefont
  {Aharonov}}, \bibinfo {author} {\bibfnamefont {L.}~\bibnamefont
  {Davidovich}}, \ and\ \bibinfo {author} {\bibfnamefont {N.}~\bibnamefont
  {Zagury}},\ }\href@noop {} {\bibfield  {journal} {\bibinfo  {journal} {Phys.
  Rev. A}\ }\textbf {\bibinfo {volume} {48}},\ \bibinfo {pages} {1687}
  (\bibinfo {year} {1993})}\BibitemShut {NoStop}%
\bibitem [{\citenamefont {Kempe}(2003)}]{kempe2003quantum}%
  \BibitemOpen
  \bibfield  {author} {\bibinfo {author} {\bibfnamefont {J.}~\bibnamefont
  {Kempe}},\ }\href@noop {} {\bibfield  {journal} {\bibinfo  {journal}
  {Contemporary Physics}\ }\textbf {\bibinfo {volume} {44}},\ \bibinfo {pages}
  {307} (\bibinfo {year} {2003})}\BibitemShut {NoStop}%
\bibitem [{\citenamefont {Nayak}\ and\ \citenamefont
  {Vishwanath}(2000)}]{nayak2000quantum}%
  \BibitemOpen
  \bibfield  {author} {\bibinfo {author} {\bibfnamefont {A.}~\bibnamefont
  {Nayak}}\ and\ \bibinfo {author} {\bibfnamefont {A.}~\bibnamefont
  {Vishwanath}},\ }\href@noop {} {\bibfield  {journal} {\bibinfo  {journal}
  {arXiv preprint quant-ph/0010117}\ } (\bibinfo {year} {2000})}\BibitemShut
  {NoStop}%
\bibitem [{\citenamefont {Chandrashekar}\ \emph {et~al.}(2008)\citenamefont
  {Chandrashekar}, \citenamefont {Srikanth},\ and\ \citenamefont
  {Laflamme}}]{chandrashekar2008optimizing}%
  \BibitemOpen
  \bibfield  {author} {\bibinfo {author} {\bibfnamefont {C.}~\bibnamefont
  {Chandrashekar}}, \bibinfo {author} {\bibfnamefont {R.}~\bibnamefont
  {Srikanth}}, \ and\ \bibinfo {author} {\bibfnamefont {R.}~\bibnamefont
  {Laflamme}},\ }\href@noop {} {\bibfield  {journal} {\bibinfo  {journal}
  {Phys. Rev. A}\ }\textbf {\bibinfo {volume} {77}},\ \bibinfo {pages} {032326}
  (\bibinfo {year} {2008})}\BibitemShut {NoStop}%
\bibitem [{\citenamefont {Zhang}\ \emph {et~al.}(2016)\citenamefont {Zhang},
  \citenamefont {Goyal}, \citenamefont {Gao}, \citenamefont {Sanders},\ and\
  \citenamefont {Simon}}]{zhang2016creating}%
  \BibitemOpen
  \bibfield  {author} {\bibinfo {author} {\bibfnamefont {W.-W.}\ \bibnamefont
  {Zhang}}, \bibinfo {author} {\bibfnamefont {S.~K.}\ \bibnamefont {Goyal}},
  \bibinfo {author} {\bibfnamefont {F.}~\bibnamefont {Gao}}, \bibinfo {author}
  {\bibfnamefont {B.~C.}\ \bibnamefont {Sanders}}, \ and\ \bibinfo {author}
  {\bibfnamefont {C.}~\bibnamefont {Simon}},\ }\href@noop {} {\bibfield
  {journal} {\bibinfo  {journal} {New Journal of Physics}\ }\textbf {\bibinfo
  {volume} {18}},\ \bibinfo {pages} {093025} (\bibinfo {year}
  {2016})}\BibitemShut {NoStop}%
\bibitem [{\citenamefont {Di~Molfetta}\ \emph {et~al.}(2014)\citenamefont
  {Di~Molfetta}, \citenamefont {Brachet},\ and\ \citenamefont
  {Debbasch}}]{di2014quantum}%
  \BibitemOpen
  \bibfield  {author} {\bibinfo {author} {\bibfnamefont {G.}~\bibnamefont
  {Di~Molfetta}}, \bibinfo {author} {\bibfnamefont {M.}~\bibnamefont
  {Brachet}}, \ and\ \bibinfo {author} {\bibfnamefont {F.}~\bibnamefont
  {Debbasch}},\ }\href@noop {} {\bibfield  {journal} {\bibinfo  {journal}
  {Physica A: Statistical Mechanics and its Applications}\ }\textbf {\bibinfo
  {volume} {397}},\ \bibinfo {pages} {157} (\bibinfo {year}
  {2014})}\BibitemShut {NoStop}%
\bibitem [{\citenamefont {Genske}\ \emph {et~al.}(2013)\citenamefont {Genske},
  \citenamefont {Alt}, \citenamefont {Steffen}, \citenamefont {Werner},
  \citenamefont {Werner}, \citenamefont {Meschede},\ and\ \citenamefont
  {Alberti}}]{genske2013electric}%
  \BibitemOpen
  \bibfield  {author} {\bibinfo {author} {\bibfnamefont {M.}~\bibnamefont
  {Genske}}, \bibinfo {author} {\bibfnamefont {W.}~\bibnamefont {Alt}},
  \bibinfo {author} {\bibfnamefont {A.}~\bibnamefont {Steffen}}, \bibinfo
  {author} {\bibfnamefont {A.~H.}\ \bibnamefont {Werner}}, \bibinfo {author}
  {\bibfnamefont {R.~F.}\ \bibnamefont {Werner}}, \bibinfo {author}
  {\bibfnamefont {D.}~\bibnamefont {Meschede}}, \ and\ \bibinfo {author}
  {\bibfnamefont {A.}~\bibnamefont {Alberti}},\ }\href@noop {} {\bibfield
  {journal} {\bibinfo  {journal} {Phys. Rev. Lett.}\ }\textbf {\bibinfo
  {volume} {110}},\ \bibinfo {pages} {190601} (\bibinfo {year}
  {2013})}\BibitemShut {NoStop}%
\bibitem [{\citenamefont {Muraleedharan}\ \emph {et~al.}(2019)\citenamefont
  {Muraleedharan}, \citenamefont {Miyake},\ and\ \citenamefont
  {Deutsch}}]{muraleedharan2019quantum}%
  \BibitemOpen
  \bibfield  {author} {\bibinfo {author} {\bibfnamefont {G.}~\bibnamefont
  {Muraleedharan}}, \bibinfo {author} {\bibfnamefont {A.}~\bibnamefont
  {Miyake}}, \ and\ \bibinfo {author} {\bibfnamefont {I.~H.}\ \bibnamefont
  {Deutsch}},\ }\href@noop {} {\bibfield  {journal} {\bibinfo  {journal} {New
  Journal of Physics}\ }\textbf {\bibinfo {volume} {21}},\ \bibinfo {pages}
  {055003} (\bibinfo {year} {2019})}\BibitemShut {NoStop}%
\bibitem [{\citenamefont {Brun}\ \emph {et~al.}(2003)\citenamefont {Brun},
  \citenamefont {Carteret},\ and\ \citenamefont
  {Ambainis}}]{PhysRevLett.91.130602}%
  \BibitemOpen
  \bibfield  {author} {\bibinfo {author} {\bibfnamefont {T.~A.}\ \bibnamefont
  {Brun}}, \bibinfo {author} {\bibfnamefont {H.~A.}\ \bibnamefont {Carteret}},
  \ and\ \bibinfo {author} {\bibfnamefont {A.}~\bibnamefont {Ambainis}},\
  }\href {\doibase 10.1103/PhysRevLett.91.130602} {\bibfield  {journal}
  {\bibinfo  {journal} {Phys. Rev. Lett.}\ }\textbf {\bibinfo {volume} {91}},\
  \bibinfo {pages} {130602} (\bibinfo {year} {2003})}\BibitemShut {NoStop}%
\bibitem [{\citenamefont {Lakshminarayan}(2003)}]{lakshminarayan2003random}%
  \BibitemOpen
  \bibfield  {author} {\bibinfo {author} {\bibfnamefont {A.}~\bibnamefont
  {Lakshminarayan}},\ }\href@noop {} {\bibfield  {journal} {\bibinfo  {journal}
  {arXiv preprint quant-ph/0305026}\ } (\bibinfo {year} {2003})}\BibitemShut
  {NoStop}%
\bibitem [{\citenamefont {Ermann}\ \emph {et~al.}(2006)\citenamefont {Ermann},
  \citenamefont {Paz},\ and\ \citenamefont {Saraceno}}]{ErmanPazSaraceno}%
  \BibitemOpen
  \bibfield  {author} {\bibinfo {author} {\bibfnamefont {L.}~\bibnamefont
  {Ermann}}, \bibinfo {author} {\bibfnamefont {J.~P.}\ \bibnamefont {Paz}}, \
  and\ \bibinfo {author} {\bibfnamefont {M.}~\bibnamefont {Saraceno}},\ }\href
  {\doibase 10.1103/PhysRevA.73.012302} {\bibfield  {journal} {\bibinfo
  {journal} {Phys. Rev. A}\ }\textbf {\bibinfo {volume} {73}},\ \bibinfo
  {pages} {012302} (\bibinfo {year} {2006})}\BibitemShut {NoStop}%
\bibitem [{\citenamefont {W\'ojcik}\ and\ \citenamefont
  {Dorfman}(2003)}]{Wojcik}%
  \BibitemOpen
  \bibfield  {author} {\bibinfo {author} {\bibfnamefont {D.~K.}\ \bibnamefont
  {W\'ojcik}}\ and\ \bibinfo {author} {\bibfnamefont {J.~R.}\ \bibnamefont
  {Dorfman}},\ }\href {\doibase 10.1103/PhysRevLett.90.230602} {\bibfield
  {journal} {\bibinfo  {journal} {Phys. Rev. Lett.}\ }\textbf {\bibinfo
  {volume} {90}},\ \bibinfo {pages} {230602} (\bibinfo {year}
  {2003})}\BibitemShut {NoStop}%
\bibitem [{\citenamefont {Kraus}\ \emph {et~al.}(1983)\citenamefont {Kraus},
  \citenamefont {B{\"o}hm}, \citenamefont {Dollard},\ and\ \citenamefont
  {Wootters}}]{kraus1983states}%
  \BibitemOpen
  \bibfield  {author} {\bibinfo {author} {\bibfnamefont {K.}~\bibnamefont
  {Kraus}}, \bibinfo {author} {\bibfnamefont {A.}~\bibnamefont {B{\"o}hm}},
  \bibinfo {author} {\bibfnamefont {J.~D.}\ \bibnamefont {Dollard}}, \ and\
  \bibinfo {author} {\bibfnamefont {W.}~\bibnamefont {Wootters}},\ }\href@noop
  {} {\bibfield  {journal} {\bibinfo  {journal} {Lecture notes in physics}\
  }\textbf {\bibinfo {volume} {190}} (\bibinfo {year} {1983})}\BibitemShut
  {NoStop}%
\bibitem [{\citenamefont {Sudarshan}\ \emph {et~al.}(1961)\citenamefont
  {Sudarshan}, \citenamefont {Mathews},\ and\ \citenamefont
  {Rau}}]{sudarshan1961stochastic}%
  \BibitemOpen
  \bibfield  {author} {\bibinfo {author} {\bibfnamefont {E.}~\bibnamefont
  {Sudarshan}}, \bibinfo {author} {\bibfnamefont {P.}~\bibnamefont {Mathews}},
  \ and\ \bibinfo {author} {\bibfnamefont {J.}~\bibnamefont {Rau}},\
  }\href@noop {} {\bibfield  {journal} {\bibinfo  {journal} {Physical Review}\
  }\textbf {\bibinfo {volume} {121}},\ \bibinfo {pages} {920} (\bibinfo {year}
  {1961})}\BibitemShut {NoStop}%
\bibitem [{\citenamefont {Konno}(2002)}]{konno2002quantum}%
  \BibitemOpen
  \bibfield  {author} {\bibinfo {author} {\bibfnamefont {N.}~\bibnamefont
  {Konno}},\ }\href@noop {} {\bibfield  {journal} {\bibinfo  {journal} {Quantum
  Information Processing}\ }\textbf {\bibinfo {volume} {1}},\ \bibinfo {pages}
  {345} (\bibinfo {year} {2002})}\BibitemShut {NoStop}%
\bibitem [{\citenamefont {Konno}(2004)}]{konno2004path}%
  \BibitemOpen
  \bibfield  {author} {\bibinfo {author} {\bibfnamefont {N.}~\bibnamefont
  {Konno}},\ }\href@noop {} {\bibfield  {journal} {\bibinfo  {journal} {arXiv
  preprint quant-ph/0406233}\ } (\bibinfo {year} {2004})}\BibitemShut {NoStop}%
\bibitem [{\citenamefont {Yang}\ \emph {et~al.}(2007)\citenamefont {Yang},
  \citenamefont {Liu},\ and\ \citenamefont {Zhang}}]{yang2007path}%
  \BibitemOpen
  \bibfield  {author} {\bibinfo {author} {\bibfnamefont {W.-S.}\ \bibnamefont
  {Yang}}, \bibinfo {author} {\bibfnamefont {C.}~\bibnamefont {Liu}}, \ and\
  \bibinfo {author} {\bibfnamefont {K.}~\bibnamefont {Zhang}},\ }\href@noop {}
  {\bibfield  {journal} {\bibinfo  {journal} {Journal of Physics A:
  Mathematical and Theoretical}\ }\textbf {\bibinfo {volume} {40}},\ \bibinfo
  {pages} {8487} (\bibinfo {year} {2007})}\BibitemShut {NoStop}%
\bibitem [{\citenamefont {Joshi}\ \emph {et~al.}(2018)\citenamefont {Joshi},
  \citenamefont {Srivatsa},\ and\ \citenamefont {Srikanth}}]{joshi2018path}%
  \BibitemOpen
  \bibfield  {author} {\bibinfo {author} {\bibfnamefont {K.~S.}\ \bibnamefont
  {Joshi}}, \bibinfo {author} {\bibfnamefont {S.}~\bibnamefont {Srivatsa}}, \
  and\ \bibinfo {author} {\bibfnamefont {R.}~\bibnamefont {Srikanth}},\
  }\href@noop {} {\bibfield  {journal} {\bibinfo  {journal} {arXiv preprint
  arXiv:1803.00448}\ } (\bibinfo {year} {2018})}\BibitemShut {NoStop}%
\bibitem [{\citenamefont {Gorin}\ \emph {et~al.}(2006)\citenamefont {Gorin},
  \citenamefont {Prosen}, \citenamefont {Seligman},\ and\ \citenamefont
  {{\v{Z}}nidari{\v{c}}}}]{gorin2006dynamics}%
  \BibitemOpen
  \bibfield  {author} {\bibinfo {author} {\bibfnamefont {T.}~\bibnamefont
  {Gorin}}, \bibinfo {author} {\bibfnamefont {T.}~\bibnamefont {Prosen}},
  \bibinfo {author} {\bibfnamefont {T.~H.}\ \bibnamefont {Seligman}}, \ and\
  \bibinfo {author} {\bibfnamefont {M.}~\bibnamefont {{\v{Z}}nidari{\v{c}}}},\
  }\href@noop {} {\bibfield  {journal} {\bibinfo  {journal} {Physics Reports}\
  }\textbf {\bibinfo {volume} {435}},\ \bibinfo {pages} {33} (\bibinfo {year}
  {2006})}\BibitemShut {NoStop}%
\bibitem [{\citenamefont {Haake}(1991)}]{haake1991quantum}%
  \BibitemOpen
  \bibfield  {author} {\bibinfo {author} {\bibfnamefont {F.}~\bibnamefont
  {Haake}},\ }in\ \href@noop {} {\emph {\bibinfo {booktitle} {Quantum Coherence
  in Mesoscopic Systems}}}\ (\bibinfo  {publisher} {Springer},\ \bibinfo {year}
  {1991})\ pp.\ \bibinfo {pages} {583--595}\BibitemShut {NoStop}%
\bibitem [{\citenamefont {Goussev}\ \emph {et~al.}(2012)\citenamefont
  {Goussev}, \citenamefont {Jalabert}, \citenamefont {Pastawski},\ and\
  \citenamefont {Wisniacki}}]{goussev2012loschmidt}%
  \BibitemOpen
  \bibfield  {author} {\bibinfo {author} {\bibfnamefont {A.}~\bibnamefont
  {Goussev}}, \bibinfo {author} {\bibfnamefont {R.~A.}\ \bibnamefont
  {Jalabert}}, \bibinfo {author} {\bibfnamefont {H.~M.}\ \bibnamefont
  {Pastawski}}, \ and\ \bibinfo {author} {\bibfnamefont {D.}~\bibnamefont
  {Wisniacki}},\ }\href@noop {} {\bibfield  {journal} {\bibinfo  {journal}
  {arXiv preprint arXiv:1206.6348}\ } (\bibinfo {year} {2012})}\BibitemShut
  {NoStop}%
\bibitem [{\citenamefont {Chandrasekhar}(1943)}]{ChandraRMP1943}%
  \BibitemOpen
  \bibfield  {author} {\bibinfo {author} {\bibfnamefont {S.}~\bibnamefont
  {Chandrasekhar}},\ }\href {\doibase 10.1103/RevModPhys.15.1} {\bibfield
  {journal} {\bibinfo  {journal} {Rev. Mod. Phys.}\ }\textbf {\bibinfo {volume}
  {15}},\ \bibinfo {pages} {1} (\bibinfo {year} {1943})}\BibitemShut {NoStop}%
\bibitem [{\citenamefont {Berry}(1979)}]{berry1979evolution}%
  \BibitemOpen
  \bibfield  {author} {\bibinfo {author} {\bibfnamefont {M.~V.}\ \bibnamefont
  {Berry}},\ }\href@noop {} {\bibfield  {journal} {\bibinfo  {journal} {Journal
  of Physics A: Mathematical and General}\ }\textbf {\bibinfo {volume} {12}},\
  \bibinfo {pages} {625} (\bibinfo {year} {1979})}\BibitemShut {NoStop}%
\bibitem [{\citenamefont {Zurek}(2001)}]{zurek2001sub}%
  \BibitemOpen
  \bibfield  {author} {\bibinfo {author} {\bibfnamefont {W.~H.}\ \bibnamefont
  {Zurek}},\ }\href@noop {} {\bibfield  {journal} {\bibinfo  {journal}
  {Nature}\ }\textbf {\bibinfo {volume} {412}},\ \bibinfo {pages} {712}
  (\bibinfo {year} {2001})}\BibitemShut {NoStop}%
\bibitem [{\citenamefont {Omanakuttan}\ and\ \citenamefont
  {Lakshminarayan}(2018)}]{omanakuttan2018quantum}%
  \BibitemOpen
  \bibfield  {author} {\bibinfo {author} {\bibfnamefont {S.}~\bibnamefont
  {Omanakuttan}}\ and\ \bibinfo {author} {\bibfnamefont {A.}~\bibnamefont
  {Lakshminarayan}},\ }\href {http://stacks.iop.org/1751-8121/51/i=38/a=385306}
  {\bibfield  {journal} {\bibinfo  {journal} {Journal of Physics A:
  Mathematical and Theoretical}\ }\textbf {\bibinfo {volume} {51}},\ \bibinfo
  {pages} {385306} (\bibinfo {year} {2018})}\BibitemShut {NoStop}%
\bibitem [{\citenamefont {Miller}\ and\ \citenamefont
  {Sarkar}(1999)}]{MillerSarkar1999}%
  \BibitemOpen
  \bibfield  {author} {\bibinfo {author} {\bibfnamefont {P.~A.}\ \bibnamefont
  {Miller}}\ and\ \bibinfo {author} {\bibfnamefont {S.}~\bibnamefont
  {Sarkar}},\ }\href {\doibase 10.1103/PhysRevE.60.1542} {\bibfield  {journal}
  {\bibinfo  {journal} {Phys. Rev. E}\ }\textbf {\bibinfo {volume} {60}},\
  \bibinfo {pages} {1542} (\bibinfo {year} {1999})}\BibitemShut {NoStop}%
\bibitem [{\citenamefont {Lakshminarayan}(2001)}]{PhysRevE.64.036207}%
  \BibitemOpen
  \bibfield  {author} {\bibinfo {author} {\bibfnamefont {A.}~\bibnamefont
  {Lakshminarayan}},\ }\href {\doibase 10.1103/PhysRevE.64.036207} {\bibfield
  {journal} {\bibinfo  {journal} {Phys. Rev. E}\ }\textbf {\bibinfo {volume}
  {64}},\ \bibinfo {pages} {036207} (\bibinfo {year} {2001})}\BibitemShut
  {NoStop}%
\bibitem [{\citenamefont {Bandyopadhyay}\ and\ \citenamefont
  {Lakshminarayan}(2002)}]{bandyopadhyay2002testing}%
  \BibitemOpen
  \bibfield  {author} {\bibinfo {author} {\bibfnamefont {J.~N.}\ \bibnamefont
  {Bandyopadhyay}}\ and\ \bibinfo {author} {\bibfnamefont {A.}~\bibnamefont
  {Lakshminarayan}},\ }\href@noop {} {\bibfield  {journal} {\bibinfo  {journal}
  {Phys. Rev. Lett.}\ }\textbf {\bibinfo {volume} {89}},\ \bibinfo {pages}
  {060402} (\bibinfo {year} {2002})}\BibitemShut {NoStop}%
\bibitem [{\citenamefont {Chaudhury}\ \emph {et~al.}(2009)\citenamefont
  {Chaudhury}, \citenamefont {Smith}, \citenamefont {Anderson}, \citenamefont
  {Ghose},\ and\ \citenamefont {Jessen}}]{Chaudhury2009}%
  \BibitemOpen
  \bibfield  {author} {\bibinfo {author} {\bibfnamefont {S.}~\bibnamefont
  {Chaudhury}}, \bibinfo {author} {\bibfnamefont {A.}~\bibnamefont {Smith}},
  \bibinfo {author} {\bibfnamefont {B.~E.}\ \bibnamefont {Anderson}}, \bibinfo
  {author} {\bibfnamefont {S.}~\bibnamefont {Ghose}}, \ and\ \bibinfo {author}
  {\bibfnamefont {P.~S.}\ \bibnamefont {Jessen}},\ }\href
  {http://www.nature.com/nature/journal/v461/n7265/suppinfo/nature08396_S1.html}
  {\bibfield  {journal} {\bibinfo  {journal} {Nature}\ }\textbf {\bibinfo
  {volume} {461}},\ \bibinfo {pages} {768} (\bibinfo {year}
  {2009})}\BibitemShut {NoStop}%
\bibitem [{\citenamefont {Neill}\ and\ \citenamefont {et~al.}(2016)}]{Neill16}%
  \BibitemOpen
  \bibfield  {author} {\bibinfo {author} {\bibfnamefont {C.}~\bibnamefont
  {Neill}}\ and\ \bibinfo {author} {\bibnamefont {et~al.}},\ }\href@noop {}
  {\bibfield  {journal} {\bibinfo  {journal} {Nat. Phys.}\ }\textbf {\bibinfo
  {volume} {12}},\ \bibinfo {pages} {1037} (\bibinfo {year}
  {2016})}\BibitemShut {NoStop}%
\bibitem [{\citenamefont {Bandyopadhyay}\ and\ \citenamefont
  {Lakshminarayan}(2004)}]{BandyoLak2004}%
  \BibitemOpen
  \bibfield  {author} {\bibinfo {author} {\bibfnamefont {J.~N.}\ \bibnamefont
  {Bandyopadhyay}}\ and\ \bibinfo {author} {\bibfnamefont {A.}~\bibnamefont
  {Lakshminarayan}},\ }\href {\doibase 10.1103/PhysRevE.69.016201} {\bibfield
  {journal} {\bibinfo  {journal} {Phys. Rev. E}\ }\textbf {\bibinfo {volume}
  {69}},\ \bibinfo {pages} {016201} (\bibinfo {year} {2004})}\BibitemShut
  {NoStop}%
\bibitem [{\citenamefont {Fujisaki}\ \emph {et~al.}(2003)\citenamefont
  {Fujisaki}, \citenamefont {Miyadera},\ and\ \citenamefont
  {Tanaka}}]{Fujisaki03}%
  \BibitemOpen
  \bibfield  {author} {\bibinfo {author} {\bibfnamefont {H.}~\bibnamefont
  {Fujisaki}}, \bibinfo {author} {\bibfnamefont {T.}~\bibnamefont {Miyadera}},
  \ and\ \bibinfo {author} {\bibfnamefont {A.}~\bibnamefont {Tanaka}},\
  }\href@noop {} {\bibfield  {journal} {\bibinfo  {journal} {Phys. Rev. E}\
  }\textbf {\bibinfo {volume} {67}},\ \bibinfo {pages} {066201} (\bibinfo
  {year} {2003})}\BibitemShut {NoStop}%
\bibitem [{\citenamefont {Pulikkottil}\ \emph {et~al.}(2020)\citenamefont
  {Pulikkottil}, \citenamefont {Lakshminarayan}, \citenamefont {Srivastava},
  \citenamefont {B{\"a}cker},\ and\ \citenamefont
  {Tomsovic}}]{pulikkottil2020entanglement}%
  \BibitemOpen
  \bibfield  {author} {\bibinfo {author} {\bibfnamefont {J.~J.}\ \bibnamefont
  {Pulikkottil}}, \bibinfo {author} {\bibfnamefont {A.}~\bibnamefont
  {Lakshminarayan}}, \bibinfo {author} {\bibfnamefont {S.~C.}\ \bibnamefont
  {Srivastava}}, \bibinfo {author} {\bibfnamefont {A.}~\bibnamefont
  {B{\"a}cker}}, \ and\ \bibinfo {author} {\bibfnamefont {S.}~\bibnamefont
  {Tomsovic}},\ }\href@noop {} {\bibfield  {journal} {\bibinfo  {journal}
  {Phys. Rev. E}\ }\textbf {\bibinfo {volume} {101}},\ \bibinfo {pages}
  {032212} (\bibinfo {year} {2020})}\BibitemShut {NoStop}%
\bibitem [{\citenamefont {Calabrese}\ and\ \citenamefont
  {Cardy}(2007)}]{Calabrese2007}%
  \BibitemOpen
  \bibfield  {author} {\bibinfo {author} {\bibfnamefont {P.}~\bibnamefont
  {Calabrese}}\ and\ \bibinfo {author} {\bibfnamefont {J.}~\bibnamefont
  {Cardy}},\ }\href {\doibase 10.1088/1742-5468/2007/10/p10004} {\bibfield
  {journal} {\bibinfo  {journal} {Journal of Statistical Mechanics: Theory and
  Experiment}\ }\textbf {\bibinfo {volume} {2007}},\ \bibinfo {pages} {P10004}
  (\bibinfo {year} {2007})}\BibitemShut {NoStop}%
\bibitem [{\citenamefont {Bardarson}\ \emph {et~al.}(2012)\citenamefont
  {Bardarson}, \citenamefont {Pollmann},\ and\ \citenamefont
  {Moore}}]{Bardarson2012}%
  \BibitemOpen
  \bibfield  {author} {\bibinfo {author} {\bibfnamefont {J.~H.}\ \bibnamefont
  {Bardarson}}, \bibinfo {author} {\bibfnamefont {F.}~\bibnamefont {Pollmann}},
  \ and\ \bibinfo {author} {\bibfnamefont {J.~E.}\ \bibnamefont {Moore}},\
  }\href {\doibase 10.1103/PhysRevLett.109.017202} {\bibfield  {journal}
  {\bibinfo  {journal} {Phys. Rev. Lett.}\ }\textbf {\bibinfo {volume} {109}},\
  \bibinfo {pages} {017202} (\bibinfo {year} {2012})}\BibitemShut {NoStop}%
\bibitem [{\citenamefont {\ifmmode \check{Z}\else
  \v{Z}\fi{}nidari\ifmmode~\check{c}\else \v{c}\fi{}}\ \emph
  {et~al.}(2008)\citenamefont {\ifmmode \check{Z}\else
  \v{Z}\fi{}nidari\ifmmode~\check{c}\else \v{c}\fi{}}, \citenamefont {Prosen},\
  and\ \citenamefont {Prelov\ifmmode~\check{s}\else
  \v{s}\fi{}ek}}]{Znidaric2008}%
  \BibitemOpen
  \bibfield  {author} {\bibinfo {author} {\bibfnamefont {M.}~\bibnamefont
  {\ifmmode \check{Z}\else \v{Z}\fi{}nidari\ifmmode~\check{c}\else
  \v{c}\fi{}}}, \bibinfo {author} {\bibfnamefont {T.~c.~v.}\ \bibnamefont
  {Prosen}}, \ and\ \bibinfo {author} {\bibfnamefont {P.}~\bibnamefont
  {Prelov\ifmmode~\check{s}\else \v{s}\fi{}ek}},\ }\href {\doibase
  10.1103/PhysRevB.77.064426} {\bibfield  {journal} {\bibinfo  {journal} {Phys.
  Rev. B}\ }\textbf {\bibinfo {volume} {77}},\ \bibinfo {pages} {064426}
  (\bibinfo {year} {2008})}\BibitemShut {NoStop}%
\bibitem [{\citenamefont {Serbyn}\ \emph {et~al.}(2013)\citenamefont {Serbyn},
  \citenamefont {Papi{\'c}},\ and\ \citenamefont
  {Abanin}}]{serbyn2013universal}%
  \BibitemOpen
  \bibfield  {author} {\bibinfo {author} {\bibfnamefont {M.}~\bibnamefont
  {Serbyn}}, \bibinfo {author} {\bibfnamefont {Z.}~\bibnamefont {Papi{\'c}}}, \
  and\ \bibinfo {author} {\bibfnamefont {D.~A.}\ \bibnamefont {Abanin}},\
  }\href@noop {} {\bibfield  {journal} {\bibinfo  {journal} {Phys. Rev. Lett.}\
  }\textbf {\bibinfo {volume} {110}},\ \bibinfo {pages} {260601} (\bibinfo
  {year} {2013})}\BibitemShut {NoStop}%
\bibitem [{\citenamefont {Park}\ and\ \citenamefont {Kim}(2003)}]{Park2003}%
  \BibitemOpen
  \bibfield  {author} {\bibinfo {author} {\bibfnamefont {H.-K.}\ \bibnamefont
  {Park}}\ and\ \bibinfo {author} {\bibfnamefont {S.~W.}\ \bibnamefont {Kim}},\
  }\href {\doibase 10.1103/PhysRevA.67.060102} {\bibfield  {journal} {\bibinfo
  {journal} {Phys. Rev. A}\ }\textbf {\bibinfo {volume} {67}},\ \bibinfo
  {pages} {060102} (\bibinfo {year} {2003})}\BibitemShut {NoStop}%
\bibitem [{\citenamefont {Nag}\ \emph {et~al.}(2001)\citenamefont {Nag},
  \citenamefont {Lahiri},\ and\ \citenamefont {Ghosh}}]{Nag2001}%
  \BibitemOpen
  \bibfield  {author} {\bibinfo {author} {\bibfnamefont {S.}~\bibnamefont
  {Nag}}, \bibinfo {author} {\bibfnamefont {A.}~\bibnamefont {Lahiri}}, \ and\
  \bibinfo {author} {\bibfnamefont {G.}~\bibnamefont {Ghosh}},\ }\href
  {\doibase https://doi.org/10.1016/S0375-9601(01)00746-0} {\bibfield
  {journal} {\bibinfo  {journal} {Phys. Rev. A}\ }\textbf {\bibinfo {volume}
  {292}},\ \bibinfo {pages} {43 } (\bibinfo {year} {2001})}\BibitemShut
  {NoStop}%
\bibitem [{\citenamefont {Zhao}\ and\ \citenamefont {Sirker}(2019)}]{Zhao2019}%
  \BibitemOpen
  \bibfield  {author} {\bibinfo {author} {\bibfnamefont {Y.}~\bibnamefont
  {Zhao}}\ and\ \bibinfo {author} {\bibfnamefont {J.}~\bibnamefont {Sirker}},\
  }\href {\doibase 10.1103/PhysRevB.100.014203} {\bibfield  {journal} {\bibinfo
   {journal} {Phys. Rev. B}\ }\textbf {\bibinfo {volume} {100}},\ \bibinfo
  {pages} {014203} (\bibinfo {year} {2019})}\BibitemShut {NoStop}%
\bibitem [{\citenamefont {Page}(1993)}]{page1993average}%
  \BibitemOpen
  \bibfield  {author} {\bibinfo {author} {\bibfnamefont {D.~N.}\ \bibnamefont
  {Page}},\ }\href@noop {} {\bibfield  {journal} {\bibinfo  {journal} {Phys.
  Rev. Lett.}\ }\textbf {\bibinfo {volume} {71}},\ \bibinfo {pages} {1291}
  (\bibinfo {year} {1993})}\BibitemShut {NoStop}%
\bibitem [{\citenamefont {Sen}(1996)}]{sen1996average}%
  \BibitemOpen
  \bibfield  {author} {\bibinfo {author} {\bibfnamefont {S.}~\bibnamefont
  {Sen}},\ }\href@noop {} {\bibfield  {journal} {\bibinfo  {journal} {Phys.
  Rev. Lett.}\ }\textbf {\bibinfo {volume} {77}},\ \bibinfo {pages} {1}
  (\bibinfo {year} {1996})}\BibitemShut {NoStop}%
\bibitem [{\citenamefont {Marcenko}\ and\ \citenamefont
  {Pastur}(1967)}]{Marcenko67}%
  \BibitemOpen
  \bibfield  {author} {\bibinfo {author} {\bibfnamefont {V.~A.}\ \bibnamefont
  {Marcenko}}\ and\ \bibinfo {author} {\bibfnamefont {L.~A.}\ \bibnamefont
  {Pastur}},\ }\href {http://stacks.iop.org/0025-5734/1/i=4/a=A01} {\bibfield
  {journal} {\bibinfo  {journal} {Mathematics of the USSR-Sbornik}\ }\textbf
  {\bibinfo {volume} {1}},\ \bibinfo {pages} {457} (\bibinfo {year}
  {1967})}\BibitemShut {NoStop}%
\bibitem [{\citenamefont {Bengtsson}\ and\ \citenamefont
  {{\.Z}yczkowski}(2017)}]{BZBook}%
  \BibitemOpen
  \bibfield  {author} {\bibinfo {author} {\bibfnamefont {I.}~\bibnamefont
  {Bengtsson}}\ and\ \bibinfo {author} {\bibfnamefont {K.}~\bibnamefont
  {{\.Z}yczkowski}},\ }\href {https://books.google.co.in/books?id=OD0yDwAAQBAJ}
  {\emph {\bibinfo {title} {Geometry of Quantum States: An Introduction to
  Quantum Entanglement}}}\ (\bibinfo  {publisher} {Cambridge University
  Press},\ \bibinfo {year} {2017})\BibitemShut {NoStop}%
\bibitem [{\citenamefont {Kubotani}\ \emph {et~al.}(2008)\citenamefont
  {Kubotani}, \citenamefont {Adachi},\ and\ \citenamefont
  {Toda}}]{Kubotani2008}%
  \BibitemOpen
  \bibfield  {author} {\bibinfo {author} {\bibfnamefont {H.}~\bibnamefont
  {Kubotani}}, \bibinfo {author} {\bibfnamefont {S.}~\bibnamefont {Adachi}}, \
  and\ \bibinfo {author} {\bibfnamefont {M.}~\bibnamefont {Toda}},\ }\href
  {\doibase 10.1103/PhysRevLett.100.240501} {\bibfield  {journal} {\bibinfo
  {journal} {Phys. Rev. Lett.}\ }\textbf {\bibinfo {volume} {100}},\ \bibinfo
  {pages} {240501} (\bibinfo {year} {2008})}\BibitemShut {NoStop}%
\bibitem [{\citenamefont {Mishra}\ and\ \citenamefont
  {Lakshminarayan}(2014)}]{Mishra2014}%
  \BibitemOpen
  \bibfield  {author} {\bibinfo {author} {\bibfnamefont {S.~K.}\ \bibnamefont
  {Mishra}}\ and\ \bibinfo {author} {\bibfnamefont {A.}~\bibnamefont
  {Lakshminarayan}},\ }\href {\doibase 10.1209/0295-5075/105/10002} {\bibfield
  {journal} {\bibinfo  {journal} {{EPL} (Europhysics Letters)}\ }\textbf
  {\bibinfo {volume} {105}},\ \bibinfo {pages} {10002} (\bibinfo {year}
  {2014})}\BibitemShut {NoStop}%
\bibitem [{\citenamefont {Pietracaprina}\ \emph {et~al.}(2017)\citenamefont
  {Pietracaprina}, \citenamefont {Parisi}, \citenamefont {Mariano},
  \citenamefont {Pascazio},\ and\ \citenamefont {Scardicchio}}]{Parisi17}%
  \BibitemOpen
  \bibfield  {author} {\bibinfo {author} {\bibfnamefont {F.}~\bibnamefont
  {Pietracaprina}}, \bibinfo {author} {\bibfnamefont {G.}~\bibnamefont
  {Parisi}}, \bibinfo {author} {\bibfnamefont {A.}~\bibnamefont {Mariano}},
  \bibinfo {author} {\bibfnamefont {S.}~\bibnamefont {Pascazio}}, \ and\
  \bibinfo {author} {\bibfnamefont {A.}~\bibnamefont {Scardicchio}},\ }\href
  {http://stacks.iop.org/1742-5468/2017/i=11/a=113102} {\bibfield  {journal}
  {\bibinfo  {journal} {Journal of Statistical Mechanics: Theory and
  Experiment}\ }\textbf {\bibinfo {volume} {2017}},\ \bibinfo {pages} {113102}
  (\bibinfo {year} {2017})}\BibitemShut {NoStop}%
\bibitem [{\citenamefont {Georgopoulos}\ \emph {et~al.}(2019)\citenamefont
  {Georgopoulos}, \citenamefont {Emary},\ and\ \citenamefont
  {Zuliani}}]{georgopoulos2019comparison}%
  \BibitemOpen
  \bibfield  {author} {\bibinfo {author} {\bibfnamefont {K.}~\bibnamefont
  {Georgopoulos}}, \bibinfo {author} {\bibfnamefont {C.}~\bibnamefont {Emary}},
  \ and\ \bibinfo {author} {\bibfnamefont {P.}~\bibnamefont {Zuliani}},\
  }\href@noop {} {\bibfield  {journal} {\bibinfo  {journal} {arXiv preprint
  arXiv:1911.00305}\ } (\bibinfo {year} {2019})}\BibitemShut {NoStop}%
\bibitem [{\citenamefont {Chen}\ \emph {et~al.}(2019)\citenamefont {Chen},
  \citenamefont {Shiau}, \citenamefont {Wu},\ and\ \citenamefont
  {Wu}}]{chen2019hybrid}%
  \BibitemOpen
  \bibfield  {author} {\bibinfo {author} {\bibfnamefont {C.-C.}\ \bibnamefont
  {Chen}}, \bibinfo {author} {\bibfnamefont {S.-Y.}\ \bibnamefont {Shiau}},
  \bibinfo {author} {\bibfnamefont {M.-F.}\ \bibnamefont {Wu}}, \ and\ \bibinfo
  {author} {\bibfnamefont {Y.-R.}\ \bibnamefont {Wu}},\ }\href@noop {}
  {\bibfield  {journal} {\bibinfo  {journal} {Scientific reports}\ }\textbf
  {\bibinfo {volume} {9}},\ \bibinfo {pages} {1} (\bibinfo {year}
  {2019})}\BibitemShut {NoStop}%
\bibitem [{\citenamefont {Volya}\ and\ \citenamefont
  {Zelevinsky}(2020)}]{volya2020time}%
  \BibitemOpen
  \bibfield  {author} {\bibinfo {author} {\bibfnamefont {A.}~\bibnamefont
  {Volya}}\ and\ \bibinfo {author} {\bibfnamefont {V.}~\bibnamefont
  {Zelevinsky}},\ }\href@noop {} {\bibfield  {journal} {\bibinfo  {journal}
  {Journal of Physics: Complexity}\ }\textbf {\bibinfo {volume} {1}},\ \bibinfo
  {pages} {025007} (\bibinfo {year} {2020})}\BibitemShut {NoStop}%
\bibitem [{\citenamefont {Yan}\ \emph {et~al.}()\citenamefont {Yan},
  \citenamefont {Cincio},\ and\ \citenamefont {Zurek}}]{yan2020information}%
  \BibitemOpen
  \bibfield  {author} {\bibinfo {author} {\bibfnamefont {B.}~\bibnamefont
  {Yan}}, \bibinfo {author} {\bibfnamefont {L.}~\bibnamefont {Cincio}}, \ and\
  \bibinfo {author} {\bibfnamefont {W.~H.}\ \bibnamefont {Zurek}},\ }\href@noop
  {} {\bibfield  {journal} {\bibinfo  {journal} {Phys. Rev. Lett.}\ }\textbf
  {\bibinfo {volume} {124}}}\BibitemShut {NoStop}%
\bibitem [{\citenamefont {Omanakuttan}\ and\ \citenamefont
  {Lakshminarayan}(2019)}]{omanakuttan2019out}%
  \BibitemOpen
  \bibfield  {author} {\bibinfo {author} {\bibfnamefont {S.}~\bibnamefont
  {Omanakuttan}}\ and\ \bibinfo {author} {\bibfnamefont {A.}~\bibnamefont
  {Lakshminarayan}},\ }\href@noop {} {\bibfield  {journal} {\bibinfo  {journal}
  {Phys. Rev. E}\ }\textbf {\bibinfo {volume} {99}},\ \bibinfo {pages} {062128}
  (\bibinfo {year} {2019})}\BibitemShut {NoStop}%
\end{thebibliography}%

\end{document}